\documentclass{mn2e}

\usepackage{graphicx}

\def\be{\begin{equation}}
\def\ee{\end{equation}}

\def\gtsima{$\; \buildrel > \over \sim \;$}
\def\ltsima{$\; \buildrel < \over \sim \;$}
\def\prosima{$\; \buildrel \propto \over \sim \;$}
\def\gsim{\lower.5ex\hbox{\gtsima}}
\def\lsim{\lower.5ex\hbox{\ltsima}}
\def\simgt{\lower.5ex\hbox{\gtsima}}
\def\simlt{\lower.5ex\hbox{\ltsima}}
\def\simpr{\lower.5ex\hbox{\prosima}}
\def\la{\lsim}
\def\ga{\gsim}

\title[Particle energy cascade in the IGM]{Particle energy cascade in the intergalactic medium}         
\author[Vald\'{e}s, Evoli \& Ferrara]
{M. Vald\'{e}s$^1$, C. Evoli$^2$ and A. Ferrara$^3$\\
$^1$ IPMU, University of Tokyo, 5-1-5 Kashiwanoha, Kashiwa, Chiba 277-8568, Japan\\
$^2$ SISSA/ISAS, via Beirut 2-4, 34151 Trieste, Italy\\
$^3$ Scuola Normale Superiore, Piazza dei Cavalieri 7, 56126 Pisa, Italy}

\begin{document}
\maketitle
\label{firstpage}

\begin{abstract}
We study the development of high energy ($E_{\rm in} \le 1$~TeV) cascades produced by a primary electron of energy $E_{\rm in}$ 
injected into the intergalactic medium (IGM). To this aim we have developed the new code MEDEA (Monte Carlo Energy 
DEposition Analysis) which includes Bremsstrahlung and Inverse Compton (IC) processes, along with H/He collisional 
ionizations and excitations, and electron-electron collisions. The cascade energy partition into heating, excitations 
and ionizations depends primarily on the IGM ionized fraction, $x_{\rm e}$, but also on redshift, $z$, due to IC on CMB  
photons. While Bremsstrahlung is unimportant under most conditions, IC becomes largely dominant at energies 
$E_{\rm in} \geq  1$ MeV. The main effect of IC at injection energies $E_{\rm in} \leq  100$ MeV is a significant boost of the
fraction of energy converted into low energy photons ($h \nu < 10.2$ eV) which do not further interact with the IGM.
For energies $E_{\rm in} \geq  1$ GeV CMB photons are preferentially upscattered within the X-ray spectrum 
($h \nu > 10^4$ eV) and can free stream to the observer. Complete tables of the fractional energy depositions 
as a function of redshift, $E_{\rm in}$ and ionized fraction are given. Our results can be used in many astrophysical contexts, 
with an obvious application related to the study of decaying/annihilating Dark Matter (DM) candidates in the high-$z$ Universe.
\end{abstract}


\section{Introduction}

High energy particles can be produced by several astrophysical and cosmological sources through
different acceleration processes. The energy stored in relativistic particles might be a non-negligible
fraction of the total energy of these systems and therefore an obvious question arises about
how this energy is eventually thermalized and transferred to the surrounding environment.
In spite of its importance, this question has received only a relatively limited attention.
Previous works have considered non-relativistic initial energies of up to 10 keV 
(e.g. Shull 1979, Shull \& van Steenberg 1985, Dalgarno et al. 1999; Vald\'{e}s \& Ferrara 2008, hereafter VF08;
Furlanetto \& Johnson Stoever 2009). In this energy 
range processes such as free-free emission with charged particles and Inverse Compton (IC) with a 
diffuse distribution of photons can be safely neglected. 

However for many astrophysical applications it is necessary to deal with higher energy particles:
extrapolating the results and fitting formulae presented in the aforementioned works can lead to substantially 
incorrect results. Particles can be accelerated to relativistic energies in a number of astrophysical systems, leading 
to the emission of synchrotron and Compton radiation observable in a broad range of energy bands. 
Active Galactic Nuclei, Stellar flares, Gamma Ray Bursts, Pulsar Wind Nebulae, Supernova Remnants
are believed to house a population of shock accelerated electrons (see e.g. MacKinnon \& Mallik 2009;   
Venter \& de Jager 2008; Resmi \& Bhattacharya 2008). It is then important to consider extensions of these works to
compute at best the evolution of the energy cascade of relativistic electrons of energy $E_{\rm in}$ into 
a partially ionized gas under realistic cosmological conditions including the presence of CMB photons. 

Radio observations of galaxy clusters have revealed the existence of the so-called radio
"relics" which can be explained with the presence in the intracluster medium of relativistic electrons 
accelerated by shocks due to cluster merging. This can give rise to X-ray emission through IC of 
CMB photons, as recent observations of the satellites Beppo-SAX and RXTE seem to indicate (see e.g. Rephaeli 1977; 
Fusco-Femiano, Landi, \& Orlandini 2007; Rephaeli et al. 2008; Ferrari 2009).

Another astrophysical source of relativistic electrons could come from the decay or annihilation
of DM particles. If indeed, as many theoretical models predict,
this elusive matter component of the Universe injects relativistic electrons and positrons into the IGM and
the consequent inverse Compton scattering with the CMB photons could generate a distortion of 
the black body spectrum by Sunyaev-Zeldovich (SZ) effect (see e.g. Lavalle, Boehm, \& Barthes 2009;
Colafrancesco \& Mele 2001; Colafrancesco 2004; Colafrancesco, Profumo, \& Ullio 2006).
In the near future the recently launched satellite Planck (The Planck collaboration, 2006) and the ALMA array (Wootten
\& Thomson 2009) could detect the SZ deviation induced by DM decays/annihilations.

In the past few years a large number of works has investigated the effects and detectability of DM decays/annihilations
into the high redshift IGM via observations of the redshifted 21~cm hyperfine triplet-singlet 
level transition of the ground state of neutral hydrogen 
(Shchekinov \& Vasiliev 2006; Furlanetto, Oh, \& Pierpaoli 2006;  Vald\'{e}s et al. 2007; 
Natarajan \& Schwarz 2009). 
The interest for such kind of studies is generated by the present or planned construction 
of large radio interferometers
such as the Low frequency Array (LOFAR, http://www.lofar.org/), the 21 Centimeter Array (21CMA, http://21cma.bao.ac.cn/), the Murchison 
Widefield Array (MWA, http://www.mwatelescope.org/) and the Square Kilometer Array (SKA, http://www.skatelescope.org/). 
These instruments will in fact 
open a new observational window by the detection 
of the HI 21 cm line at $z>6$ (e.g. Peterson, Pen \& Wu 2005; Bowman, Morales, 
\& Hewitt 2005; Kassim et al. 2004; Wyithe, Loeb, \& Barnes 2005).

To understand if observations of the redshifted HI 21 cm line can help constrain DM it is crucial 
to follow in detail the 
energy cascade from energetic primary photons or electrons up to energies much higher than previously 
studied (Shull \& van Steenberg 1985, VF08) to include those DM candidates that can produce relativistic electrons. 
In our calculations we choose an initial electron energy 1 MeV $< E_{\rm in} < 1$ TeV.
The details of the energy degradation of electrons in
this energy range are of great interest for many of the aforementioned astrophysical sources
and in particular for the study of the effects of DM decays/annihilations, since recent 
data from PAMELA, ATIC, FERMI-LAT and HESS experiments have shown that there
are electron-positron excesses in the cosmic ray energy spectrum which can be explained 
by annihilations of DM between 1 to 2 TeV (see e.g. Cirelli, Franceschini, \& Strumia 2008;
Liu et al. 2009; Berg et al. 2009; Hooper \& Tait 2009;
Chen, Takahashi \& Yanagida 2009; He 2009 for a brief review with an extensive list of citations).
Recently Slatyer et al. (2009) computed the energy injection rates for some annihilating DM 
candidates.

The rest of the paper is organized as follows. In Sec. 2 we describe the relevant physical process
and their implementation into the multi-purpose code MEDEA (Monte Carlo Energy DEposition Analysis) we have developed 
to follow the energy cascade. In Sec. 3 we present the results and analyze their dependence on redshift, 
initial electron energy and redshift. In Sec. 4 we discuss the implications and draw some conclusions.
The complete numerical outputs of MEDEA can be found in tabulated form in Appendix A.

\section{Method}

Our code MEDEA, a substantial extension of the algorithm described in VF08, 
is based on a Monte Carlo scheme that allows to follow the energy cascade arising 
from the interaction of relativistic electrons
($1$ MeV $<E_{\rm in}<1$ TeV) with the IGM for $10<z<50$.

A Monte Carlo method is a computational algorithm that relies on repeated 
random sampling of the relevant physical quantities and processes (e.g. cross-sections and
interaction probabilities) to follow the evolution of the system.
Essentially the code calculates for every particle the probability of the main
interaction channels and then selects one by a random number generator.
Once the reaction happens the code follows the resulting particles to the next interaction,
until the energy of the particle drops below a given threshold taken in our case to be 10.2 
eV (the Lyman $\alpha$ transition energy), when the photon-gas interaction rate vanishes
(in the absence of heavy elements, molecules or dust). 

For each assumed primary energy $E_{\rm in}$, gas ionized fraction $x_{\rm e}$, and redshift, $z$, we 
performed 100 Monte Carlo realizations. Although in VF08 we performed 1000 realizations for 
each parameter choice, the full MEDEA proved to be very computationally expensive since 
the large energies considered imply a much more complex cascade. In particular, IC interactions 
with the CMB required the code to follow in detail a large amount of upscattered photons.
In VF08 we found that while 1000 Monte Carlo realizations gave highly consistent results, 
going from 1000 to 50 realizations changed the averaged values by less than $5\%$ and the respective 
$\sigma$ by less than $10\%$. Therefore performing 100 realization per parameter choice does not 
bring any substantial bias due to the random nature of the computation.

In this Section the physics related to the low energy processes $E_{\rm in} \le 10^4$eV will be
reviewed briefly and defer the reader to VF08 for details. We will instead describe extensively 
the implementation of the two newly included high energy processes, Bremsstrahlung and Inverse Compton, 
which were previously not included or negligible.

\subsection{Low energy regime}

MEDEA follows in detail the fate of relativistic electrons by calculating
the energy cascade that originates from a number of interactions with the surrounding gas.
When a secondary electron or photon of energy 
below $10$ keV is produced the code behaves consistently to the calculation described in VF08.
An ionizing photon of energy $h \nu <10$ keV will be immediately converted into an electron since 
the dominant interaction in this case is photoionization of an H or He atom (see e.g.
Zdziarski \& Svensson 1989). 
Then the code follows the fate of each of these electrons by calculating the 
mean free paths relative to a number of possible interactions: H, He, HeI 
collisional ionizations;  H, He collisional excitations; collisions with thermal 
electrons; free-free interactions with
ionized atoms; recombinations. The latter two are virtually negligible as stated in VF08.
We assume that $x_{\rm e} \equiv n\mbox{(H}^+\mbox{)}/n\mbox{(H)}\equiv n\mbox{(He}^+\mbox{)}/n\mbox{(He)} $
and that the helium fraction by mass is $f_{\rm He}=$ 0.248 in agreement with the 3-yr Wilkinson 
Microwave Anisotropy Probe (WMAP) data analysis (Spergel et al. 2007).

Every collisional ionization event by an electron with an H or He atom generates another free electron
whose further interactions with the gas will be followed in detail by the code. These energy ramifications find 
an end when secondary electron energies drops below 10.2 eV and their entire energy is 
subsequently deposited as heat by thermalization with gas. This is a good assumption at the redshifts of
interest and in general for a gas with a temperature below $10^4$~K.
The energy distribution of electrons generated after collisional ionization of H and He are treated as in VF08.

We also follow individual photons emitted by excited atoms returning to the ground state
after a collision with a free electron. Photons with energies lower than 10.2 eV do not interact 
further with the gas while higher energy photons do and are followed by the code.
As in VF08 also transitions to/from the $2s$ level of HI are considered, i.e. direct collisional 
excitations to $2s$ followed by the two-photon forbidden transition $2s \rightarrow 1s$
or indirect cascades from $n\geq $ 3 states which happen preferentially
through the $2s$ level rather than through $2p$ (see e.g. Hirata 2006, Chuzhoy \& Shapiro 2007).

Electron-electron collisional cross section resulting in an energy loss $\Delta E_{\rm in}$ is given by:
\begin{equation}\label{}
\sigma_{\rm ee} = 40 \pi e^4 \ln{\Lambda} \left( \frac{0.05}{f}\right) E_{\rm in}^{-2} \mbox{cm}^2
\end{equation} 
where the Coulomb logarithm $\ln{\Lambda}=\ln{(4E_{\rm in}/ \zeta_{\rm e})}$ with $\zeta_{\rm e}=7.40\times 10^{-11} (n_{\rm e}/{\rm cm}^{-3})$ eV 
(Furlanetto \& Johnson Stoever 2009) 
and $f= \Delta E_{\rm in} / E_{\rm in}=0.05$ is chosen
to account for the discrete nature of the calculation.
This process is now implemented with higher accuracy with respect to VF08 and this eliminates the
discrepancy in the heating fractional energy deposition 
$f_{\rm h}$ at moderate and high ionized fractions observed by Furlanetto \& Johnson Stoever (2009).

After Coulomb scattering events kinetic energy is transferred elastically to the thermal 
electrons present in the IGM and our code follows consistently the fate of the upscattered
electrons.

We report here briefly the references for the cross sections used to account for the several processes 
considered: (i) collisional ionization of H, He, He+ were taken from Kim \& Rudd (1994), Shah et al. (1987), 
Shah et al. (1988); collisional excitation cross sections of H and He are from Stone, Kim \& Desclaux (2002);
collisional excitation cross section to the $2s$ level of H is from Bransden \& Noble (1976);
cross section for Coulomb collisions between electrons is from Spitzer \& Scott (1969).

The hydrogen recombination cross section is given by
\begin{equation}\label{}
\sigma_{\rm r}(\nu,n)\approx 3 \times 10^{10}\frac{g_{\rm fb}}{\nu n^3 v_{\rm e}^2}\,\, \mbox{cm}^2
\end{equation}
where $g_{\rm fb}$ is the Gaunt factor of $\cal O$(1), calculated and tabulated by Karzas \& Latter (1961),
$\nu$ is the emitted radiation frequency, $v_{\rm e}$ is the electron velocity 
and $n$ is the level at which the electron recombines. 

\subsection{Bremsstrahlung}

In a free-free (or Bremsstrahlung) interaction between an electrons with either a proton, ionized He atom or neutral 
H and He atoms the electron decelerates and therefore a continuum photon is emitted.
This process has been studied in great detail by several groups and the theoretical 
and experimental results gathered are assembled in a few review works  
(see e.g. Koch \& Motz 1959, Blumenthal \& Gould 1970) where analytical cross section formulae 
are given for a number of different conditions and approximations.

In general the exact Bremsstrahlung cross sections can be derived by the methods of quantum electrodynamics
but it is possible to make reasonable assumptions and simplifications
to obtain approximated cross sections by a semiclassical approach.
We refer the reader to the aforementioned references for details on the approximated cross sections.

\begin{figure}
\centering
\includegraphics[width=6.5cm,angle=270]{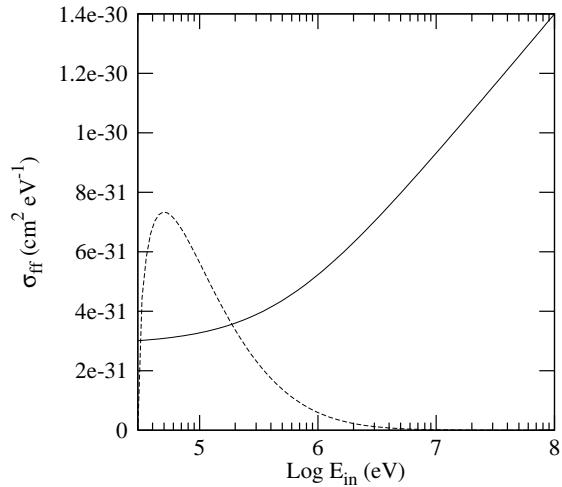}
\caption{\small Cross section for Bremsstrahlung leading to the emission of a 30 keV photon 
as a function of the initial electron kinetic energy. The two curves represent the non-relativistic 
(dashed line) and relativistic limit (solid line). \label{graph1}}
\end{figure}

\begin{figure}
\centering
\includegraphics[width=6.5cm,angle=270]{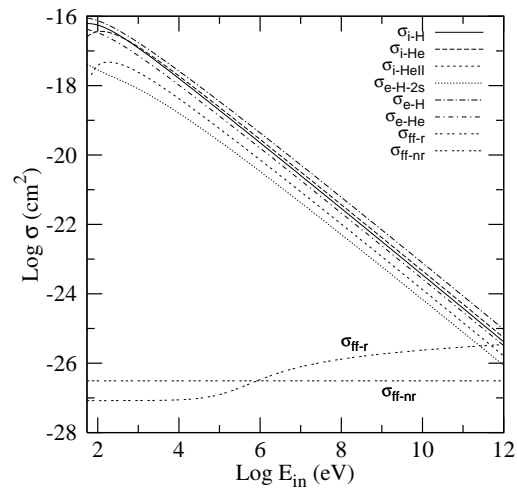}
\caption{\small Comparison between interactions included in MEDEA.
The labels within the plot box from top to bottom stand respectively for the following cross sections: 
H ionization; He ionization; HeII ionization; H excitation to the 2s level; H excitation;
He excitation; relativistic free-free; non-relativistic free-free.\label{graph1}}
\end{figure}

The two main regimes to consider are the relativistic and non-relativistic one.
In Fig 1 we show the cross section for electron-proton Bremsstrahlung leading to the emission of a 30 keV photon 
as a function of the initial electron kinetic energy in the non-relativistic and relativistic limits.
Bremsstrahlung can also take place between free electrons and atoms, in which case 
screening by atomic electrons has to be taken into account in the calculations.

We include in our calculation two different cross sections for the non-relativistic and relativistic case. 
We find that for most of the considered energy range Bremsstrahlung is negligible in comparison to other processes.
This can be seen clearly in Fig 2 which represent a direct comparison of the cross sections of
the several processes included in MEDEA. 
We do not include inverse Compton in the plot since it is related to CMB photons and 
it is therefore not visually comparable to the other interactions.
The two curves labeled $\sigma_{\rm ff-nr}$ and $\sigma_{\rm ff-r}$ in the plot are the total radiation cross section derived
by the Born approximation procedure for the cases: 
\begin{itemize}
\item Non-relativistic; not screened
\item Relativistic; not screened
\end{itemize} 
respectively. Only for values $E_{\rm in}\geq 10^{11}$ eV $\sigma_{\rm ff-r}$ becomes comparable 
to the other cross sections reported.

\subsection{High energy regime: Inverse Compton}

When electron energies become relativistic Inverse Compton with CMB photons becomes by far dominant.
To understand why we first introduce briefly the physics of this particular process.
The following is the inverse Compton angle averaged cross section in 
$\mbox{cm}^{-2}\,\mbox{eV}^{-1}$.

\begin{figure*}
   \centering 
   \includegraphics[width=8cm]{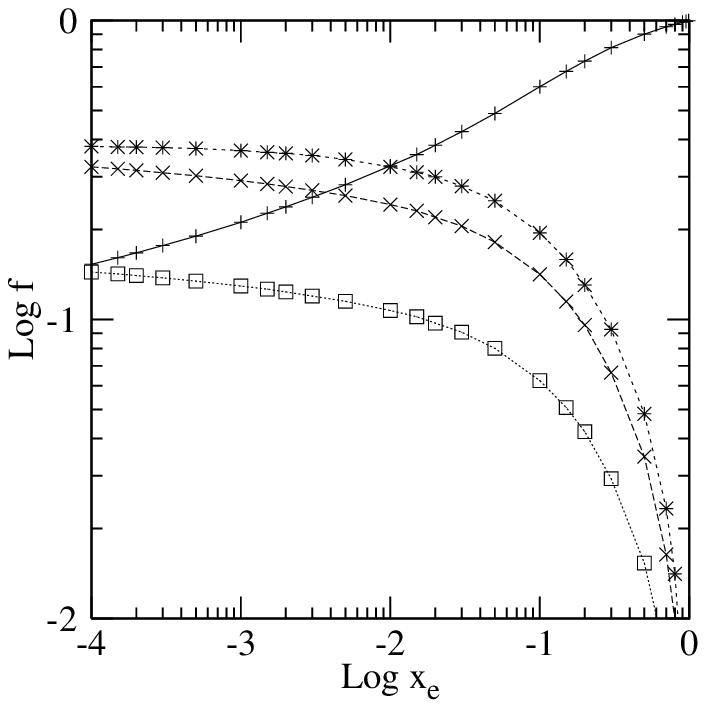}
   \includegraphics[width=8cm]{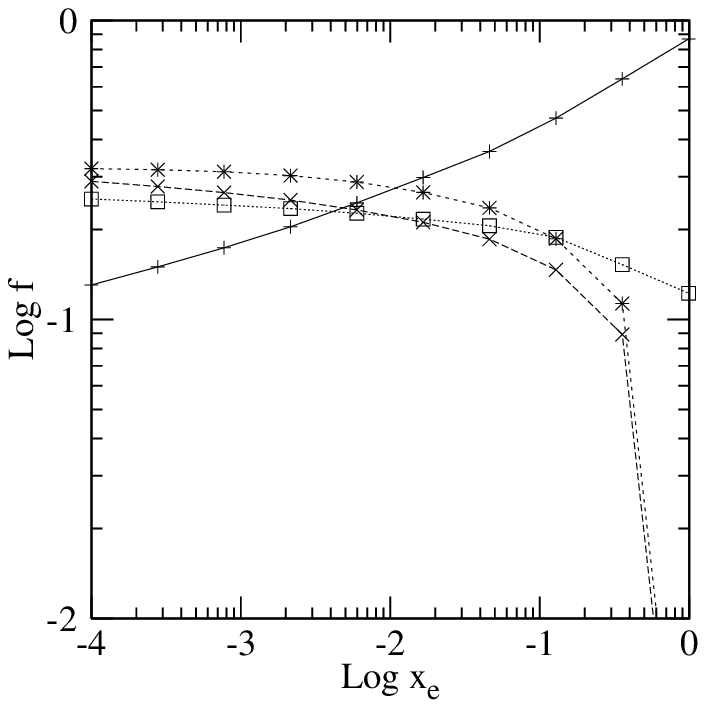}
   \caption{{\it Left panel}: Fractional energy losses for a primary 10 keV electron. The data points stand for: 
photons with $E <$ 10.2eV (\textit{squares}); gas heating (\textit{pluses}); Ly$\alpha$ 
photons (\textit{crosses}), and ionizations (\textit{asterisks}). The calculation is performed for 25 values of $x_{\rm e}$
chosen between $x_{\rm e}=0.0001$ and $x_{\rm e}=0.99$. 
{\it Right}: Same for the case of a 1 MeV primary electron. The fractional energy depositions in this case are calculated
for 9 values of of $x_{\rm e}$ chosen between $x_{\rm e}=0.0001$ and $x_{\rm e}=0.99$.\label{fig:1}}
\end{figure*} 

\begin{equation}
\sigma_{\rm KN}(E_{\rm in}^{\gamma}, E_{\rm fin}^{\gamma}, \gamma_{\rm e})=\frac{3 \sigma_{\rm T}}{4 {\gamma_{\rm e}}^2 E_{\rm in}^{\gamma}} G(q, \eta).
\end{equation}
We denote here by $E_{\rm in}^{\gamma}$, $E_{\rm fin}^{\gamma}$ and ${\gamma_{\rm e}}$ the energy of the incoming and outgoing photon 
and the Lorentz factor of the incoming electron, respectively. $\sigma_T= 8\pi {r_0}^2/3 $ is the Thomson cross-section
while the function $G(q, \eta)$ is given by:

\begin{equation}
G(q, \eta)=2q \ln{q}+(1+2q)(1-q)+2 \eta q(1-q),
\end{equation}

\begin{equation}
q=\frac{E_{\rm fin}^{\gamma}}{\Gamma (\gamma_{\rm e} m_{\rm e} c^2 - E_{\rm fin}^{\gamma})}, \,\,\,\,\,\, \Gamma = \frac{4 E_{\rm in}^{\gamma} \gamma_{\rm e}}{m_{\rm e} c^2}, \,\,\,\,\,\, \eta = \frac{E_{\rm in}^{\gamma} E_{\rm fin}^{\gamma}}{(m_{\rm e} c^2)^2}.
\end{equation}
The energy distribution of the scattered photons is:

\begin{eqnarray}
\frac{dN_{\rm \gamma, E_{\rm in}^{\gamma}}}{dt d E_{\rm fin}^{\gamma}}&=&\frac{2 \pi {r_0}^2 c}{{\gamma_{\rm e}}^2} \frac{n(E_{\rm in}^{\gamma})dE_{\rm in}^{\gamma}}{E_{\rm in}^{\gamma}} G(q, \eta)= \nonumber\\
                                                         & & \nonumber \\
                                                          & & = c \sigma_{\rm KN} n(E_{\rm in}^{\gamma})dE_{\rm in}^{\gamma}.
\end{eqnarray}
The only assumption for the validity of these equations is that $\gamma_{\rm e} \gg 1$, i.e. that the electron is relativistic.
If the energy of the incoming photon in the rest frame of the electron ${E_{\rm in}^{\gamma \,\, *}} = \gamma_{\rm e} E_{\rm in}^{\gamma} (1 - \beta \cos{\theta})\ll m_{\rm e} c^2$ 
we are in the Thomson limit, therefore $\Gamma \ll 1$ and the 
last term in $G(q, \eta)$ becomes negligible. The opposite case ${E_{\rm in}^{\gamma \,\, *}}\gg m_{\rm e} c^2$ is the extreme Klein-Nishina limit.
If we consider a CMB photon at $z\sim 10$ ($E_{\rm in}^{\gamma}\sim 2.35 \times 10^{-3}$ eV) and electron energies up to 1 TeV ($\gamma_{\rm e} \sim 10^6$)
we are still well within the Thomson limit; we will therefore assume that scatterings happen in this regime
throughout the rest of this work.
Kinematic requirements in the Thomson limit set a lower and upper limit on the energy of the up-scattered CMB photon (see e.g.
Blumenthal \& Gould 1970; Petruk 2009):

\begin{equation}
E_{\rm fin(min)}^{\gamma}= E_{\rm in}^{\gamma}, \,\,\,\,\, E_{\rm fin(max)}^{\gamma} = 4 {\gamma_{\rm e}}^2 E_{\rm in}^{\gamma}
\end{equation}
For our purposes we consider 
$\sigma_{\rm KN}(E_{\rm in}^{\gamma}, E_{\rm fin}^{\gamma}, \gamma_{\rm e})$
which is given in cm$^2$ eV$^{-1}$:
from this quantity we can derive the total inverse-Compton cross-section for an electron and a photon with 
assigned initial energies by integrating eq. (2) on the energy of the outgoing photon,

\begin{equation}
\sigma_{\rm IC}(E_{\rm in}^{\gamma}, \gamma_{\rm e})= \int_{\rm E_{\rm in}^{\gamma}}^{4 {\gamma_{\rm e}}^2 E_{\rm in}^{\gamma}} {d E_{\rm fin}^{\gamma} \frac{3 \sigma_T}{4 {\gamma_{\rm e}}^2 E_{\rm in}^{\gamma}} G(E_{\rm fin}^{\gamma})}.
\end{equation}
We can now derive the probability for a given high energy electron to scatter off CMB photons 
described by the distribution $n(E_{\rm in}^{\gamma})$
As in VF08 we compute the mean free path for inverse Compton to compare it with those from the
other processes. 
We have that, for the generic i-th process, $\lambda_i=1/ \sigma_i n_i$, therefore:

\begin{equation}
\lambda_{\rm IC}(\gamma_{\rm e})= \frac{1}{ \int{d E_{\rm in}^{\gamma} n(E_{\rm in}^{\gamma}) \sigma_{\rm IC}(E_{\rm in}^{\gamma}, \gamma_{\rm e})}}.
\end{equation}
This quantity is dependent only on the initial electron energy if the isotropic initial photon energy distribution is known.
Now we can calculate the relative probability of having inverse Compton and can follow the route of the electron
into the IGM by Monte Carlo technique.

If the electron actually scatters off a CMB photon we need to compute the energy kick to the photon and therefore
the energy loss of the electron.
To do so we have to determine first which is the initial photon energy since we computed the mean free path in a 
distribution of photons. For this purpose we have to use the rejection method to generate sampling photons 
from the distribution given by $n(E_{\rm in}^{\gamma})\cdot \sigma_{\rm IC}(E_{\rm in}^{\gamma}, \gamma_{\rm e})$.

Once we determine $E_{\rm in}^{\gamma}$ we can use similarly the rejection method on the distribution
$\sigma_{\rm KN}(E_{\rm in}^{\gamma}, E_{\rm fin}^{\gamma}, \gamma_{\rm e})$ to infer the outgoing photon energy - and therefore the electron's
energy loss.

Because we operate in the Thomson limit the mean free path of an electron in a photon background is 
very well approximated by:

\begin{equation}
\lambda_{\rm IC}= \frac{1}{\sigma_{\rm T} \, \int{d E_{\rm in}^{\gamma} n(E_{\rm in}^{\gamma}) }}.
\end{equation}
Inverse Compton is by far the process that requires the higher computational effort 
when following the cascade evolution since (i) it is the most probable interaction for a wide range of 
energies and (ii) energy loss by upscattering of CMB photons generally happens 
thorough a very high number of interactions that lower the electron kinetic 
energy by a small fraction at each step, accordingly to the limits given by Eq. (7).

In agreement with the analytical comparison 
between the collisional ionization process and inverse Compton by Hansen \& Haiman (2004) we find
that the latter is the dominant energy loss process at energies

\begin{equation}
E_{\rm in}=\gamma_{\rm e} m_{\rm e} c^2> \left(\frac{1+z}{21} \right)^{-1/2} \mbox{MeV}.
\end{equation}

\section{Results}

We present our results in Figs. 3-6 and report the tabulated fractional energy depositions in 
Appendix A, Tables A1-A21, for $z=10$, $30$, 50, for values of the ionized
fraction $x_{\rm e}$ regularly spaced in log-scale with values ranging from $10^{-4}$ to $0.99$ and for energies
between 1 MeV and 1 TeV.
The columns indicate, from left to right, which fraction of the initial electron energy is deposited into 
heat, Ly$\alpha$ excitations, ionizations, photons with $E <$ 10.2 eV, photons with $E>10^4$ eV and
the total energy of CMB photons before they are upscattered. Hereafter we will refer 
to these values as $f_{\rm h}$, $f_{\rm a}$, $f_{\rm i}$, $f_{\rm c}$, $f_{\rm HE}$ and $f_{\rm CMB}$ respectively. 
The latter column is just a test required to ensure that energy is conserved.

\begin{figure*}
\centering
\includegraphics[width=14cm, viewport=0 -130 480 310,clip]{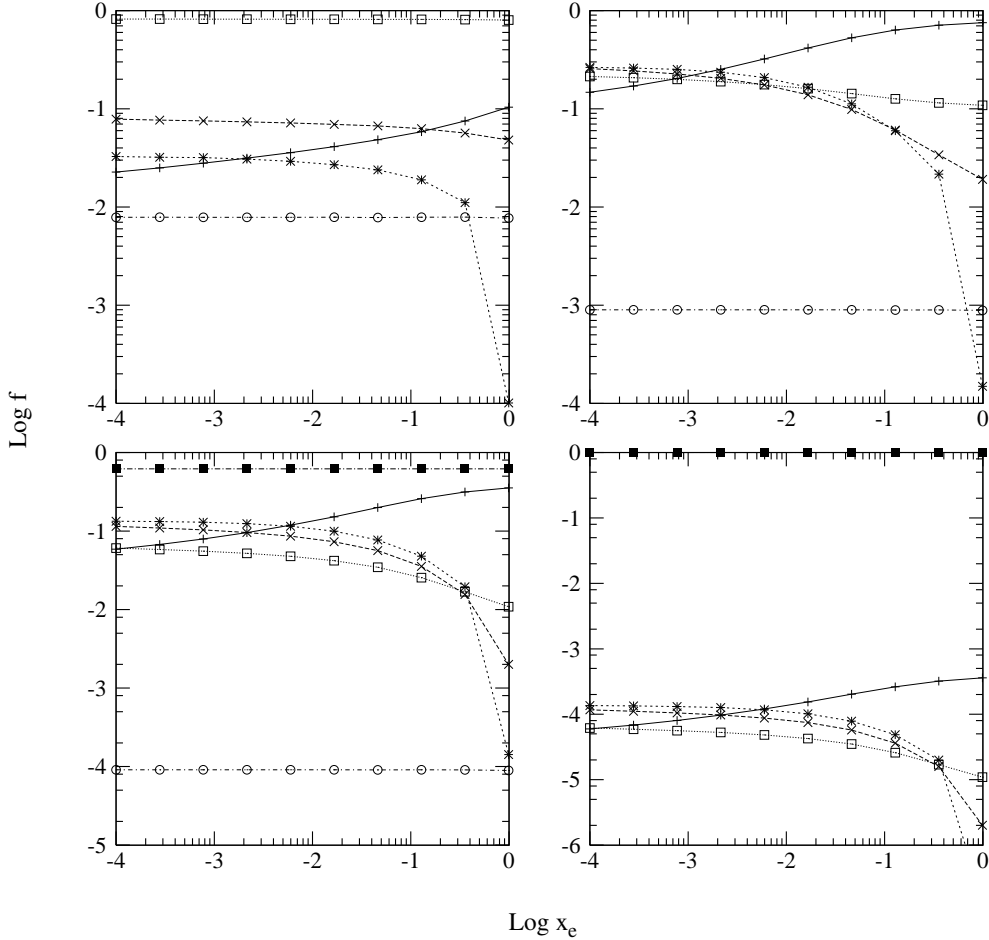}
\caption{\small Fractional energy losses for a primary electron of energy 10 MeV, 100 MeV, 
1 GeV, 1 TeV, left to right, top to bottom panels. The data points stand for: 
photons with $E <$ 10.2eV (\textit{squares}); gas heating (\textit{pluses}); Ly$\alpha$ 
photons (\textit{crosses}); ionizations (\textit{asterisks}); high energy photons with $h \nu > 10^4$ eV 
(\textit{filled squares}); energy of upscattered CMB photons (\textit{circles}).\label{graph1}}
\end{figure*}

The results, consistently with VF08, depend on the ionized fraction $x_{\rm e}$ which remains 
the most important parameter for the calculations. We also find a less marked dependence on $z$
due to the implementation in MEDEA of IC interactions with CMB photons.  

In Fig 3 we show the differences between the 10 keV results described in VF08 and the lowest energy
case considered here, $E_{\rm in}=$ 1 MeV. Notice that in the latter we perform the calculation
for only 9 different choices of $x_{\rm e}$ because of the 
higher computational burden required by running MEDEA. The considered redshift is $z=10$.
While $f_{\rm i}$, $f_{\rm a}$ and $f_{\rm h}$ appear to have similar behaviors in the two cases it is evident that
$f_{\rm c}$ is increased in the 1 MeV plot.
The reason for this is the inclusion of IC: even at relatively low energies in fact IC 
cross section is dominating. The range of energies of the upscattered CMB photons is
however very narrow (see eq. 7) and therefore individual events will enhance the photon energy to values 
$0.00259 \, \mbox{eV} \leq  h \nu \leq  0.0905$ eV. We assumed here for simplicity that the CMB photon
is at the mean CMB energy at $z=10$ but in the code we include the proper CMB spectral distribution.

\begin{figure*}
\centering
\includegraphics[width=14cm, viewport=-250 -35 230 405,clip]{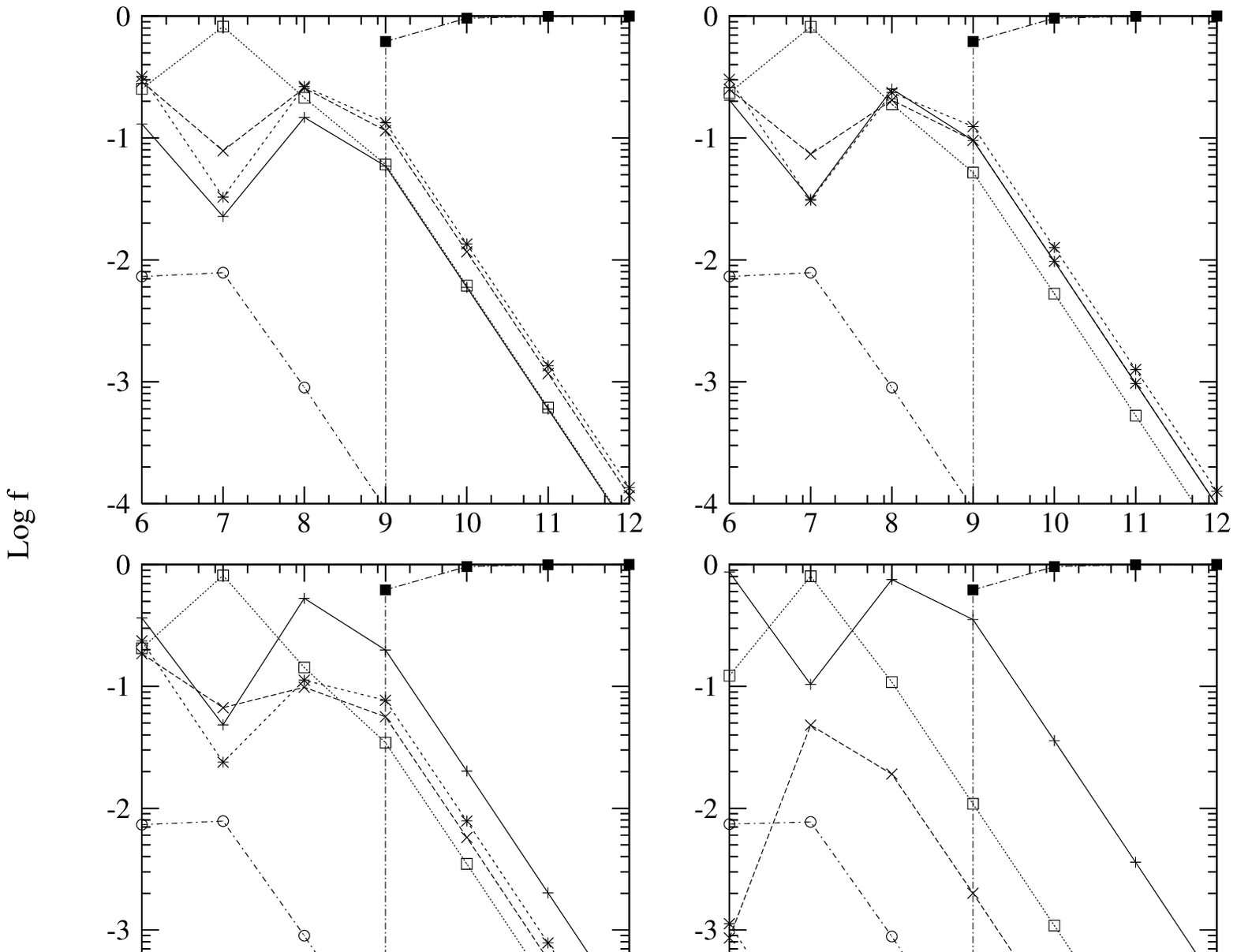}
\caption{\small Fractional energy losses as a function of initial electron energy at $z=10$ for  
$x_{\rm e}=1.000\cdot 10^{-4}, \, 2.147\cdot 10^{-3}, \, 4.610\cdot 10^{-2}, \, 9.900\cdot 10^{-1}$ 
left to right, top to bottom panels.  The data points as in Fig 4 stand for: 
photons with $E <$ 10.2eV (\textit{squares}); gas heating (\textit{pluses}); Ly$\alpha$ 
photons (\textit{crosses}); ionizations (\textit{asterisks}); high energy photons with $h \nu > 10^4$ eV 
(\textit{filled squares}); energy of upscattered CMB photons (\textit{circles}).\label{graph1}}
\end{figure*}
 
So many CMB photons are upscattered that even though the energy injection from the electrons 
is small the overall effect is still a significant increase of $f_{\rm c}$, by a factor $\sim 2$ for low
values of $x_{\rm e}$ and by over an order of magnitude for $x_{\rm e}=0.99$.
The reason for the different rise of the curve for low and high values of $x_{\rm e}$ is simply that IC
is not affected by $x_{\rm e}$, therefore the curve increases by a fixed value $\sim 0.12$ with respect to the 10 keV case.

\begin{figure*}
\centering
\includegraphics[width=12cm]{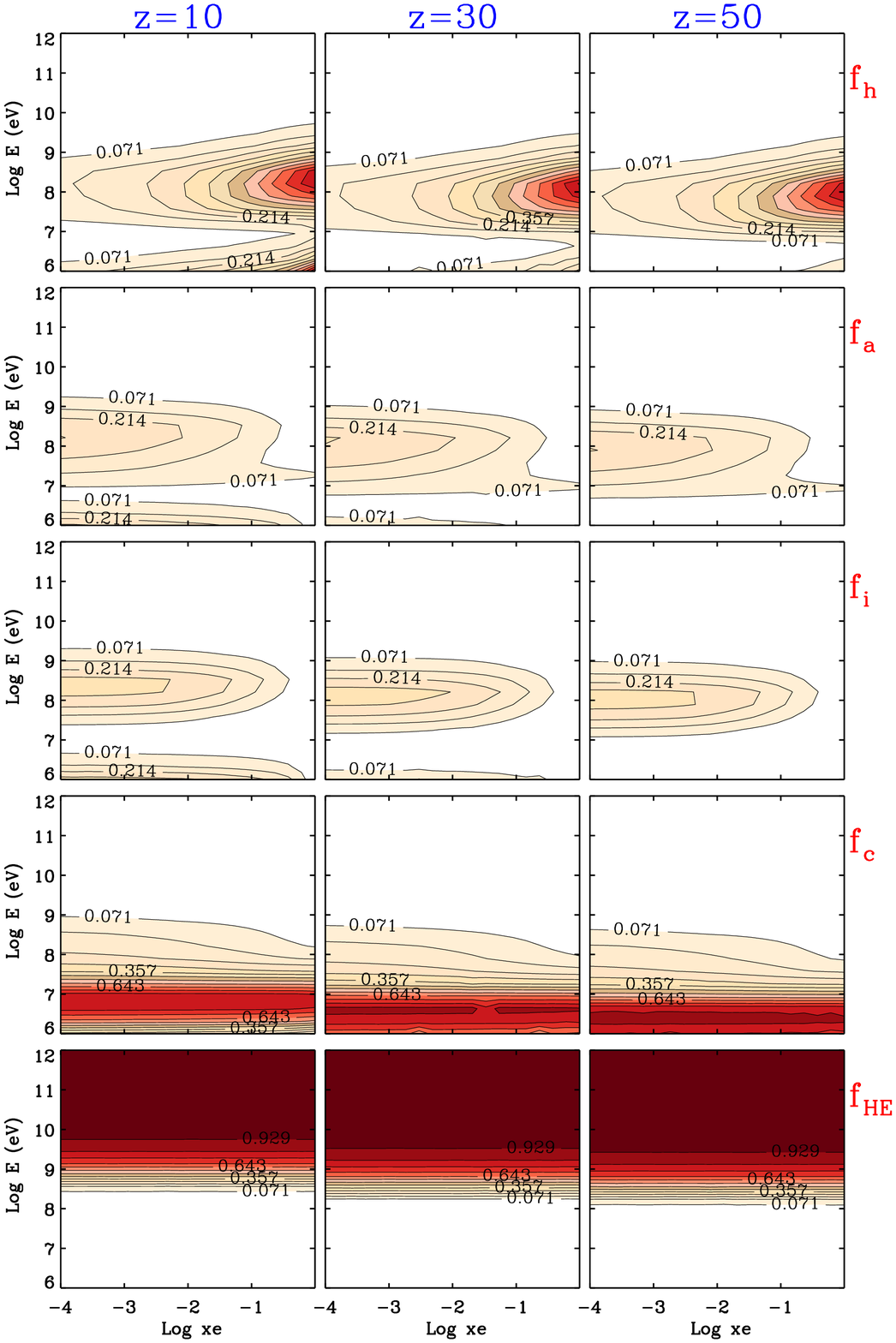}
\caption{\small Isocontour plots of the fractional energy depositions 
as a function of $E_{\rm in}$ and $x_{\rm e}$. The panels from top to bottom
are relative to $f_{\rm h}$, $f_{\rm a}$, $f_{\rm i}$, $f_{\rm c}$, $f_{\rm HE}$: these are 
the fraction of the initial electron energy that is deposited into 
heat, Ly$\alpha$ excitations, ionizations, photons with $E <$ 10.2 eV, photons with $E>10^4$ eV respectively.\label{graph1}}
\end{figure*}

When the primary electron energy is degraded by the numerous IC scatterings to values below 10 keV 
then the secondary cascade behaves consistently with the results shown in the left panel of Fig.3 and therefore the 
fractional energy depositions with the exception of $f_{\rm c}$ retain the same ratios relative to each other.

Notice that the process other than IC that produces continuum photons at energies lower than 10.2 eV is, as mention earlier, 
the two photon forbidden transition $2s \rightarrow 1s$ which we included in VF08 and that
was neglected in previous studies (e.g. Shull \& van Steenberg 1985).

We focus now on the higher energy results presented in Fig 4. Here the panels show the fractional energy depositions for
energies 10 MeV (top left panel), 100 MeV (top right), 1 GeV (bottom left) and 1 TeV (bottom right).
Two new kind of data points appear in some of the panels: the filled squares represent the fraction of the total
energy that is deposited into high energy electrons with $h \nu > 10^4$ eV ($f_{\rm HE}$) while the circles show the 
integrated energy of CMB photons involved in an IC scatter ($f_{\rm CMB}$), $before$ they are actually scattered. We keep track 
of the latter for an energy conservation sanity check.

When a photon is upscattered by IC there are four main energy ranges that we treat differently. Photons with energies 
below 10.2 eV are added to $f_{\rm c}$; the rare photons with energies between 10.2 eV and 13.6 eV (which we denote
as Lyman-continuum photons) are converted into Ly$\alpha$ photons (therefore increasing $f_{\rm a}$) while the difference
in energy is added to $f_{\rm c}$; photons with $13.6 \, \mbox{eV} \leq  h \nu \leq  10^4$ eV are assumed 
to ionize an atom and are converted into free electrons which we keep following in detail; photons 
with $h \nu > 10^4$ eV instead free stream into the IGM and are added up to the fraction $f_{\rm HE}$
(see e.g. Zdziarski \& Svensson 1989; Slatyer et al. 2009).
Obviously the range of energies will depend on the energy of the primary electron, as the maximum
energy of the upscattered photon is proportional to the square of the Lorentz factor or the electron $\gamma_{\rm e}$.

The differences that we identified when going from the 10 keV to the 1 MeV case are sharply enhanced if 
we consider a higher initial electron energy $E_{\rm in}=$ 10 MeV.
Inverse Compton remains in fact dominant but the maximum energy that the electron can give to an average 
$z=10$ CMB photon is now of the order of 5 eV.
This means that there is a dramatic boost in the fractional deposition $f_{\rm c}$, which reaches 
an almost constant value of $\sim 0.8$. The other fractional energy curves are left unchanged relative to each other
except for $f_{\rm a}$ which is raised by some CMB photons with higher than average energy that are
upscattered to Lyman-continuum values.

The 100 MeV panel shows curves with very similar properties to the 1 MeV case.
The reason for this is that now IC can upscatter photons up to values of a few hundred of eV, which 
ionize atoms, are converted into electrons and behave, consistently with VF08, as in the left panel of Fig.3.

A major difference in behavior is reported in the panel corresponding to 1 GeV. For such high energy primary
electrons the IC scattering events preferentially increase CMB photon energies to values $h \nu > 10$ keV.
In this case the $f_{\rm HE}$ curve makes its first appearance and is already dominant, with a very flat 
behavior with value $\sim $ 0.61.
The other curves are left practically unchanged relative to each other, but are lowered
to $\sim 30\%$ of the values calculated for the 10 keV case.

Finally the last panel reports the curves for a 1 TeV primary electron. For such a high initial energy
$f_{\rm HE}$=0.99 is entirely dominant, while the other fractional energy depositions are reduced 
to around $1\%$ of the 1 MeV values.
Even in this case the amount of $E_{\rm in}$ that goes into ionizations, excitations and heating 
is hardly negligible: for a 1 TeV electron there is still a total energy of $\sim 10^{-4}\times 10^{12} = 100$ MeV 
that is injected into the surrounding gas.

In Fig 5 we plot the fractional energy depositions (same symbols correspond to the same quantities
as in Fig 4) but as a function of the initial electron kinetic energy and for fixed values of the
ionized fraction, $x_{\rm e}=1.000\cdot 10^{-4}, \, 2.147\cdot 10^{-3}, \, 4.610\cdot 10^{-2}, \, 9.900\cdot 10^{-1}$
for panels left to right top to bottom. Redshift is fixed at $z=10$.

In the panels it is possible to identify the effects of IC at different energies.
For instance the bump in $f_{\rm c}$ for primary electron energy $10^7$ eV is clearly visible in all the panels.
The sharp decrease of $f_{\rm CMB}$ is due to the fact that as electrons become relativistic IC encounters 
upscatter CMB photons to high energies. Therefore the electron will need less interactions to loose its 
kinetic energy.

On the other hand at around 1 GeV there is a sharp increase in the IC high energy photons and 
therefore the other curves will decrease accordingly. At 1 TeV $f_{\rm HE}\sim 1$.

Fig 6 is a summary of our results as it reports the isocontour plot of the 
fractional energy depositions as a function of both 
the ionized fraction $x_{\rm e}$ and the initial electron energy $E_{\rm in}$.
One interesting feature is visible in the panels relative to the fractional energy that goes
into ionizations, $f_{\rm i}$. At $z=$ 10 there is a clear double peak, with the values decreasing sharply
for $E_{\rm in}\sim 10$ MeV. This is a behavior that we reported as we commented the first panel in Fig. 4
corresponding to an initial energy of 10 MeV: IC is already dominant but is unable to upscatter
CMB photons to energies higher than 10.2 eV. These photons, described by $f_{\rm c}$, do not interact further with the
IGM and therefore almost 80\% of the initial electron energy is lost and 
the values for the remaining fractional energy depositions decrease sharply.
As soon as IC preferentially upscatters photons to energies over 13.6 eV $f_{\rm i}$ rises again.
Notice that at $z=$ 50 this double peak effect is not present because CMB photons are more energetic and therefore
$f_{\rm c}$ (second line of panels from the bottom) in the range $1$ MeV $<E_{\rm in}<10$ MeV remains higher 
with respect to the lower $z$ cases.

It is also worth noticing that $f_{\rm c}$ and $f_{\rm HE}$ are essentially independent from $x_{\rm e}$
but vary slowly with redshift. Intuitively at higher redshift, when CMB photons have higher energies, the
value of $f_{\rm HE}$ grows faster with increasing $E_{\rm in}$. 

In the Appendix A we present the results for a number of MEDEA runs with different parameter choices.
The tables report the fractional energy depositions as a function of $x_{\rm e}$ given a fixed $z$ and $E_{\rm in}$.
Because of the large amount of output data we do not present a complete list of fitting formulae
but we refer the reader interested in a particular application to use the following convenient functional 
forms as in VF08: 

\begin{itemize}
\item Fraction $f_{\rm h}$ of the primary energy deposited into heat:
\begin{equation}\label{}
f_{\rm h}= 1.0-a\,(\,1.0-x^{b})
\end{equation}
\item Fraction $f_{\rm \alpha}$ of the primary energy converted into Ly$\alpha$ radiation:
\begin{equation}\label{}
f_{\rm \alpha}= a\,(\,1.0-x^{b})^{c}
\end{equation}
\item Fraction $f_{\rm i}$ of the primary energy deposited into ionizations:
\begin{equation}\label{}
f_{\rm i}= a\,(\,1.0 - x^{b})^{c}
\end{equation}
\item Fraction $f_{\rm c}$ of the primary energy deposited into continuum radiation ($h \nu < 10.2$ eV):
\begin{equation}\label{}
f_{\rm c}= a\,(\,1.0 - x^{b})
\end{equation}
\end{itemize} 
where all free parameters depend on the redshift and can be fixed by comparing the 
functions to their tabulated values in Appendix A. 

The fractional energy deposition $f_{\rm HE}$ essentially does not depend on $x_{\rm e}$ 
and can be assumed to be constant, while $f_{\rm CMB}$ is only reported 
as a check of energy conservation in every run.

Notice that the errors become smaller for high values of $E_{\rm in}$ until eventually reaching 
zero in some cases. This is just a natural effect of the calculation and value zero means that the 
uncertainty becomes smaller than the last significant digit reported in the tables.
 
\section{Conclusions}

We have introduced our code MEDEA, based on a Monte Carlo scheme that makes possible to follow the fate of 
electrons of energies up to 1 TeV in their secondary energy cascade. Our results represent a substantial
generalization of previous works that considered exclusively non-relativistic electrons.

To perform our calculation we implemented in the code a large number of interactions such as 
collisional ionizations of H, He, HeI; collisional excitations of H, He; 
electron-electron Coulomb scattering; free-free interactions of electrons with protons; 
IC with CMB photons; direct collisional excitations to the $2s$ level of HI; indirect cascades 
from $n\geq 3$ states of HI through the $2s$ level; recombinations.
The energy range of the primary electron is $1$ MeV $<E_{\rm in}<10$ MeV, the ionized fraction 
considered is $10^{-4}<x_{\rm e}<0.99$ and redshift spans $10<z<50$.

The results presented here and summarized in numerical form in the Appendix Tables, can be used for many 
astrophysical applications such as cluster radio relics, DM decays/annihilations, Active Galactic Nuclei, 
Stellar flares, Gamma Ray Bursts, Pulsar Wind Nebulae, Supernova Remnants
and, more generally, whenever it is necessary to deal with the interaction of energetic and/or
relativistic particles with the surrounding thermal gas. 

We will present new results and updates at the URL {\small http://www.arcetri.astro.it/twiki/bin/view/DAVID/MedeaCode} periodically.

\section*{Acknowledgments}

We thank Naoki Yoshida for his useful suggestions and Steven Furlanetto for insightful comments.
This work was supported in part by the Japan Society for the Promotion of Science
(JSPS) Postdoctoral Fellowship
For Foreign Researchers, Grant-in-Aid for Scientific Research (08804) and by 
the World Premier International Research Center Initiative (WPI Program).

\appendix

\section[]{Tabulated results \\
for z=10, 30, 50}

In this appendix we present the tabulated results from several of our MEDEA runs in which we fix $z$ and $E_{\rm in}$
and vary the values of the ionized fraction $x_{\rm e}$ from $10^{-4}$ to 0.99. 

\begin{table*}\label{table1}
  \begin{center}
    \begin{tabular}{|c|c|c|c|c|c|c|}\hline
      \hline 
       $x_{\rm e}$ (ionized   & Gas     &  Excitations     & Ionizations      &      Excitations  &    HE photons  & Energy from \\
      fraction)          & Heating & (Lyman-$\alpha$) & (H, He, HeII)    & ($E<10.2$ eV)   &   ($E> 10$ keV)   &   CMB\\
      \hline 
1.000e-04 &1.3042e-01$\pm$5e-04 &2.9008e-01$\pm$7e-04 &3.1945e-01$\pm$8e-04 &2.5296e-01$\pm$7e-04 &0.0000e-00$\pm$0e-00 &7.3151e-03$\pm$1e-05\\
2.779e-04 &1.5004e-01$\pm$4e-04 &2.7835e-01$\pm$2e-03 &3.1702e-01$\pm$2e-03 &2.4749e-01$\pm$2e-04 &0.0000e-00$\pm$0e-00 &7.2790e-03$\pm$7e-05\\
7.725e-04 &1.7376e-01$\pm$1e-03 &2.6586e-01$\pm$2e-03 &3.1202e-01$\pm$1e-03 &2.4126e-01$\pm$1e-03 &0.0000e-00$\pm$0e-00 &7.2688e-03$\pm$6e-05\\
2.147e-03 &2.0427e-01$\pm$1e-03 &2.5082e-01$\pm$1e-03 &3.0280e-01$\pm$1e-03 &2.3493e-01$\pm$1e-03 &0.0000e-00$\pm$0e-00 &7.3067e-03$\pm$2e-05\\
5.968e-03 &2.4545e-01$\pm$5e-04 &2.3235e-01$\pm$1e-03 &2.8824e-01$\pm$1e-03 &2.2678e-01$\pm$6e-04 &0.0000e-00$\pm$0e-00 &7.2973e-03$\pm$1e-05\\
1.658e-02 &2.9845e-01$\pm$2e-03 &2.1149e-01$\pm$1e-03 &2.6623e-01$\pm$1e-03 &2.1657e-01$\pm$9e-04 &0.0000e-00$\pm$0e-00 &7.3044e-03$\pm$1e-05\\
4.610e-02 &3.6502e-01$\pm$1e-03 &1.8574e-01$\pm$8e-04 &2.3621e-01$\pm$1e-03 &2.0574e-01$\pm$9e-04 &0.0000e-00$\pm$0e-00 &7.3117e-03$\pm$4e-05\\
1.281e-01 &4.7142e-01$\pm$8e-03 &1.4661e-01$\pm$3e-03 &1.8631e-01$\pm$3e-03 &1.8834e-01$\pm$1e-03 &0.0000e-00$\pm$0e-00 &7.3315e-03$\pm$0e-00\\
3.561e-01 &6.3845e-01$\pm$2e-02 &8.9128e-02$\pm$9e-03 &1.1313e-01$\pm$1e-02 &1.5249e-01$\pm$1e-02 &0.0000e-00$\pm$0e-00 &6.7952e-03$\pm$1e-03\\
9.900e-01 &8.6832e-01$\pm$1e-04 &8.5926e-04$\pm$7e-05 &1.1260e-03$\pm$7e-05 &1.2230e-01$\pm$4e-05 &0.0000e-00$\pm$0e-00 &7.3883e-03$\pm$0e-00\\
     \hline       
    \end{tabular}
  \end{center}
  \caption{Fraction of the energy $E_{\rm in}$ of a 1 MeV primary electron that is deposited into heat, ionizations, Ly$\alpha$ line radiation, 
photons with energy $E <$ 10.2 eV, photons with energies $E > 10$ keV due to Inverse Compton. The last column shows the total energy from the CMB 
photons before they were upscattered as a test of energy conservation. We consider here redshift $z=10$.} 
\end{table*}

\begin{table*}\label{table1}
  \begin{center}
    \begin{tabular}{|c|c|c|c|c|c|c|}\hline
      \hline 
       $x_{\rm e}$ (ionized   & Gas     &  Excitations     & Ionizations      &      Excitations  &    HE photons  & Energy from \\
      fraction)          & Heating & (Lyman-$\alpha$) & (H, He, HeII)    & ($E<10.2$ eV)   &   ($E> 10$ keV)   &   CMB\\
      \hline 
1.000e-04 &2.2719e-02$\pm$2e-04 &7.8486e-02$\pm$4e-04 &3.2693e-02$\pm$6e-04 &8.1969e-01$\pm$1e-03 &0.0000e-00$\pm$0e-00 &7.8553e-03$\pm$3e-05\\
2.779e-04 &2.5005e-02$\pm$1e-04 &7.6755e-02$\pm$2e-04 &3.2122e-02$\pm$1e-04 &8.1969e-01$\pm$3e-04 &0.0000e-00$\pm$0e-00 &7.8756e-03$\pm$0e-00\\
7.725e-04 &2.7974e-02$\pm$1e-04 &7.5307e-02$\pm$4e-04 &3.1826e-02$\pm$1e-04 &8.1854e-01$\pm$5e-04 &0.0000e-00$\pm$0e-00 &7.8607e-03$\pm$1e-05\\
2.147e-03 &3.1428e-02$\pm$1e-04 &7.3635e-02$\pm$2e-04 &3.0723e-02$\pm$1e-04 &8.1775e-01$\pm$2e-04 &0.0000e-00$\pm$0e-00 &7.8636e-03$\pm$1e-05\\
5.968e-03 &3.5781e-02$\pm$5e-04 &7.1812e-02$\pm$4e-04 &2.9210e-02$\pm$2e-04 &8.1690e-01$\pm$1e-03 &0.0000e-00$\pm$0e-00 &7.8469e-03$\pm$5e-05\\
1.658e-02 &4.1173e-02$\pm$1e-04 &6.9448e-02$\pm$9e-05 &2.6939e-02$\pm$5e-05 &8.1597e-01$\pm$1e-04 &0.0000e-00$\pm$0e-00 &7.8704e-03$\pm$0e-00\\
4.610e-02 &4.8401e-02$\pm$9e-04 &6.6941e-02$\pm$5e-04 &2.3878e-02$\pm$4e-04 &8.1438e-01$\pm$1e-03 &0.0000e-00$\pm$0e-00 &7.8214e-03$\pm$1e-04\\
1.281e-01 &5.8480e-02$\pm$7e-04 &6.2816e-02$\pm$5e-04 &1.8917e-02$\pm$2e-04 &8.1332e-01$\pm$2e-04 &0.0000e-00$\pm$0e-00 &7.8741e-03$\pm$0e-00\\
3.561e-01 &7.5130e-02$\pm$1e-03 &5.6736e-02$\pm$6e-04 &1.1091e-02$\pm$6e-04 &8.1061e-01$\pm$5e-04 &0.0000e-00$\pm$0e-00 &7.8853e-03$\pm$0e-00\\
9.900e-01 &1.0380e-01$\pm$8e-03 &4.8136e-02$\pm$2e-04 &1.0091e-04$\pm$1e-05 &8.0172e-01$\pm$8e-03 &0.0000e-00$\pm$0e-00 &7.7063e-03$\pm$2e-04\\
      \hline
    \end{tabular}
  \end{center}
  \caption{Same as table A1 but for energy $E_{\rm in}=$10 MeV.} 
\end{table*}

\begin{table*}\label{table1}
  \begin{center}
    \begin{tabular}{|c|c|c|c|c|c|c|}\hline
      \hline 
       $x_{\rm e}$ (ionized   & Gas     &  Excitations     & Ionizations      &      Excitations  &    HE photons  & Energy from \\
      fraction)          & Heating & (Lyman-$\alpha$) & (H, He, HeII)    & ($E<10.2$ eV)   &   ($E> 10$ keV)   &   CMB\\
      \hline 
1.000e-04 &1.4748e-01$\pm$7e-05 &2.5673e-01$\pm$1e-04 &2.6408e-01$\pm$1e-04 &2.1349e-01$\pm$8e-05 &0.0000e-00$\pm$0e-00 &8.9852e-04$\pm$0e-00\\
2.779e-04 &1.7047e-01$\pm$5e-05 &2.4354e-01$\pm$1e-04 &2.6018e-01$\pm$1e-04 &2.0753e-01$\pm$9e-05 &0.0000e-00$\pm$0e-00 &8.9666e-04$\pm$0e-00\\
7.725e-04 &2.0363e-01$\pm$8e-05 &2.2647e-01$\pm$1e-04 &2.5193e-01$\pm$1e-04 &1.9978e-01$\pm$1e-04 &0.0000e-00$\pm$0e-00 &8.9668e-04$\pm$0e-00\\
2.147e-03 &2.5220e-01$\pm$1e-04 &2.0427e-01$\pm$2e-04 &2.3621e-01$\pm$2e-04 &1.8928e-01$\pm$5e-04 &0.0000e-00$\pm$0e-00 &8.9712e-04$\pm$0e-00\\
5.968e-03 &3.2200e-01$\pm$1e-04 &1.7515e-01$\pm$1e-04 &2.0818e-01$\pm$1e-04 &1.7646e-01$\pm$8e-05 &0.0000e-00$\pm$0e-00 &8.9806e-04$\pm$0e-00\\
1.658e-02 &4.1623e-01$\pm$1e-04 &1.3907e-01$\pm$1e-04 &1.6585e-01$\pm$1e-04 &1.6051e-01$\pm$6e-05 &0.0000e-00$\pm$0e-00 &8.9863e-04$\pm$0e-00\\
4.610e-02 &5.2806e-01$\pm$7e-05 &9.8462e-02$\pm$8e-05 &1.1235e-01$\pm$8e-05 &1.4282e-01$\pm$8e-05 &0.0000e-00$\pm$0e-00 &8.9781e-04$\pm$0e-00\\
1.281e-01 &6.3468e-01$\pm$1e-04 &6.0682e-02$\pm$5e-05 &5.9844e-02$\pm$6e-05 &1.2647e-01$\pm$2e-04 &0.0000e-00$\pm$0e-00 &8.9363e-04$\pm$1e-05\\
3.561e-01 &7.1102e-01$\pm$4e-05 &3.3984e-02$\pm$8e-05 &2.1616e-02$\pm$6e-05 &1.1500e-01$\pm$1e-04 &0.0000e-00$\pm$0e-00 &8.9426e-04$\pm$1e-05\\
9.900e-01 &7.5398e-01$\pm$3e-04 &1.9103e-02$\pm$3e-05 &1.4946e-04$\pm$0e-00 &1.0848e-01$\pm$4e-04 &0.0000e-00$\pm$0e-00 &8.8537e-04$\pm$2e-05\\
      \hline
    \end{tabular}
  \end{center}
  \caption{Same as table A1 but for energy $E_{\rm in}=$100 MeV.} 
\end{table*}

\begin{table*}\label{table1}
  \begin{center}
    \begin{tabular}{|c|c|c|c|c|c|c|}\hline
      \hline 
       $x_{\rm e}$ (ionized   & Gas     &  Excitations     & Ionizations      &      Excitations  &    HE photons  & Energy from \\
      fraction)          & Heating & (Lyman-$\alpha$) & (H, He, HeII)    & ($E<10.2$ eV)   &   ($E> 10$ keV)   &   CMB\\
      \hline 
1.000e-04 &5.8745e-02$\pm$1e-04 &1.1476e-01$\pm$2e-04 &1.3340e-01$\pm$3e-04 &6.0638e-02$\pm$1e-04 &6.1871e-01$\pm$8e-04 &9.0828e-05$\pm$0e-00\\
2.779e-04 &6.7436e-02$\pm$2e-04 &1.0939e-01$\pm$3e-04 &1.3194e-01$\pm$4e-04 &5.8250e-02$\pm$1e-04 &6.1924e-01$\pm$1e-03 &9.1010e-05$\pm$0e-00\\
7.725e-04 &7.9248e-02$\pm$1e-04 &1.0337e-01$\pm$2e-04 &1.2943e-01$\pm$3e-04 &5.5519e-02$\pm$1e-04 &6.1867e-01$\pm$8e-04 &9.1025e-05$\pm$0e-00\\
2.147e-03 &9.5438e-02$\pm$1e-04 &9.5723e-02$\pm$1e-04 &1.2410e-01$\pm$1e-04 &5.2018e-02$\pm$6e-05 &6.1898e-01$\pm$5e-04 &9.1020e-05$\pm$0e-00\\
5.968e-03 &1.1865e-01$\pm$2e-04 &8.6119e-02$\pm$1e-04 &1.1495e-01$\pm$1e-04 &4.7679e-02$\pm$6e-05 &6.1883e-01$\pm$6e-04 &9.0949e-05$\pm$0e-00\\
1.658e-02 &1.5150e-01$\pm$2e-04 &7.3118e-02$\pm$1e-04 &9.9394e-02$\pm$2e-04 &4.1908e-02$\pm$6e-05 &6.2029e-01$\pm$7e-04 &9.0998e-05$\pm$0e-00\\
4.610e-02 &1.9905e-01$\pm$4e-04 &5.6280e-02$\pm$1e-04 &7.6884e-02$\pm$2e-04 &3.4537e-02$\pm$8e-05 &6.1945e-01$\pm$9e-04 &9.0639e-05$\pm$0e-00\\
1.281e-01 &2.5848e-01$\pm$4e-04 &3.5535e-02$\pm$8e-05 &4.7792e-02$\pm$1e-04 &2.5526e-02$\pm$3e-05 &6.1886e-01$\pm$6e-04 &9.1055e-05$\pm$0e-00\\
3.561e-01 &3.1456e-01$\pm$7e-04 &1.5581e-02$\pm$4e-05 &1.9543e-02$\pm$6e-05 &1.6829e-02$\pm$3e-05 &6.1967e-01$\pm$8e-04 &9.0561e-05$\pm$0e-00\\
9.900e-01 &3.5383e-01$\pm$5e-04 &2.0031e-03$\pm$0e-00 &1.4197e-04$\pm$0e-00 &1.0883e-02$\pm$1e-05 &6.1933e-01$\pm$5e-04 &8.9216e-05$\pm$0e-00\\
      \hline
    \end{tabular}
  \end{center}
  \caption{Same as table A1 but for energy $E_{\rm in}=$1 GeV.} 
\end{table*}

\begin{table*}\label{table1}
  \begin{center}
    \begin{tabular}{|c|c|c|c|c|c|c|}\hline
      \hline 
       $x_{\rm e}$ (ionized   & Gas     &  Excitations     & Ionizations      &      Excitations  &    HE photons  & Energy from \\
      fraction)          & Heating & (Lyman-$\alpha$) & (H, He, HeII)    & ($E<10.2$ eV)   &   ($E> 10$ keV)   &   CMB\\
      \hline 
1.000e-04 &5.9462e-03$\pm$2e-05 &1.1624e-02$\pm$4e-05 &1.3514e-02$\pm$5e-05 &6.1291e-03$\pm$2e-05 &9.6141e-01$\pm$1e-04 &9.1150e-06$\pm$0e-00\\
2.779e-04 &6.8388e-03$\pm$2e-05 &1.1103e-02$\pm$3e-05 &1.3394e-02$\pm$4e-05 &5.8970e-03$\pm$1e-05 &9.6139e-01$\pm$1e-04 &9.0818e-06$\pm$0e-00\\
7.725e-04 &8.0239e-03$\pm$1e-05 &1.0471e-02$\pm$2e-05 &1.3119e-02$\pm$3e-05 &5.6112e-03$\pm$1e-05 &9.6140e-01$\pm$9e-05 &9.1103e-06$\pm$0e-00\\
2.147e-03 &9.6907e-03$\pm$2e-05 &9.7353e-03$\pm$2e-05 &1.2629e-02$\pm$4e-05 &5.2745e-03$\pm$1e-05 &9.6129e-01$\pm$1e-04 &9.1165e-06$\pm$0e-00\\
5.968e-03 &1.2006e-02$\pm$1e-05 &8.7330e-03$\pm$1e-05 &1.1658e-02$\pm$1e-05 &4.8212e-03$\pm$0e-00 &9.6140e-01$\pm$4e-05 &9.1141e-06$\pm$0e-00\\
1.658e-02 &1.5348e-02$\pm$2e-05 &7.4298e-03$\pm$1e-05 &1.0100e-02$\pm$1e-05 &4.2429e-03$\pm$0e-00 &9.6150e-01$\pm$6e-05 &9.1052e-06$\pm$0e-00\\
4.610e-02 &2.0207e-02$\pm$3e-05 &5.7406e-03$\pm$1e-05 &7.8432e-03$\pm$1e-05 &3.4864e-03$\pm$3e-05 &9.6134e-01$\pm$6e-05 &9.2516e-06$\pm$0e-00\\
1.281e-01 &2.6227e-02$\pm$8e-05 &3.6150e-03$\pm$1e-05 &4.8683e-03$\pm$2e-05 &2.5793e-03$\pm$0e-00 &9.6132e-01$\pm$1e-04 &9.1122e-06$\pm$0e-00\\
3.561e-01 &3.1949e-02$\pm$6e-05 &1.5881e-03$\pm$0e-00 &1.9956e-03$\pm$0e-00 &1.6912e-03$\pm$0e-00 &9.6139e-01$\pm$7e-05 &9.0471e-06$\pm$0e-00\\
9.900e-01 &3.5922e-02$\pm$5e-05 &2.0074e-04$\pm$0e-00 &1.4512e-05$\pm$0e-00 &1.0864e-03$\pm$0e-00 &9.6139e-01$\pm$4e-05 &9.0136e-06$\pm$0e-00\\
      \hline
    \end{tabular}
  \end{center}
  \caption{Same as table A1 but for energy $E_{\rm in}=$10 GeV.} 
\end{table*}

\begin{table*}\label{table1}
  \begin{center}
    \begin{tabular}{|c|c|c|c|c|c|c|}\hline
      \hline 
       $x_{\rm e}$ (ionized   & Gas     &  Excitations     & Ionizations      &      Excitations  &    HE photons  & Energy from \\
      fraction)          & Heating & (Lyman-$\alpha$) & (H, He, HeII)    & ($E<10.2$ eV)   &   ($E> 10$ keV)   &   CMB\\
      \hline 
1.000e-04 &5.9639e-04$\pm$0e-00 &1.1654e-03$\pm$0e-00 &1.3556e-03$\pm$0e-00 &6.1424e-04$\pm$0e-00 &9.9613e-01$\pm$0e-00 &9.0855e-07$\pm$0e-00\\
2.779e-04 &6.8470e-04$\pm$0e-00 &1.1117e-03$\pm$0e-00 &1.3414e-03$\pm$0e-00 &5.9045e-04$\pm$0e-00 &9.9613e-01$\pm$0e-00 &9.1100e-07$\pm$0e-00\\
7.725e-04 &8.0140e-04$\pm$0e-00 &1.0457e-03$\pm$0e-00 &1.3100e-03$\pm$0e-00 &5.6057e-04$\pm$0e-00 &9.9614e-01$\pm$0e-00 &9.1199e-07$\pm$0e-00\\
2.147e-03 &9.6794e-04$\pm$0e-00 &9.7279e-04$\pm$0e-00 &1.2610e-03$\pm$0e-00 &5.2704e-04$\pm$0e-00 &9.9613e-01$\pm$0e-00 &9.1070e-07$\pm$0e-00\\
5.968e-03 &1.2015e-03$\pm$0e-00 &8.7412e-04$\pm$0e-00 &1.1669e-03$\pm$0e-00 &4.8255e-04$\pm$0e-00 &9.9613e-01$\pm$0e-00 &9.1039e-07$\pm$0e-00\\
1.658e-02 &1.5374e-03$\pm$0e-00 &7.4403e-04$\pm$0e-00 &1.0119e-03$\pm$0e-00 &4.2480e-04$\pm$0e-00 &9.9614e-01$\pm$0e-00 &9.1118e-07$\pm$0e-00\\
4.610e-02 &2.0177e-03$\pm$0e-00 &5.7279e-04$\pm$0e-00 &7.8245e-04$\pm$0e-00 &3.4990e-04$\pm$0e-00 &9.9613e-01$\pm$1e-05 &9.1113e-07$\pm$0e-00\\
1.281e-01 &2.6192e-03$\pm$0e-00 &3.6098e-04$\pm$0e-00 &4.8611e-04$\pm$0e-00 &2.5767e-04$\pm$0e-00 &9.9613e-01$\pm$0e-00 &9.0617e-07$\pm$0e-00\\
3.561e-01 &3.1890e-03$\pm$1e-05 &1.5844e-04$\pm$0e-00 &1.9910e-04$\pm$0e-00 &1.6954e-04$\pm$0e-00 &9.9614e-01$\pm$1e-05 &9.1139e-07$\pm$0e-00\\
9.900e-01 &3.5979e-03$\pm$0e-00 &2.0092e-05$\pm$0e-00 &1.4512e-06$\pm$0e-00 &1.0892e-04$\pm$0e-00 &9.9613e-01$\pm$0e-00 &8.9809e-07$\pm$0e-00\\
      \hline
    \end{tabular}
  \end{center}
  \caption{Same as table A1 but for energy $E_{\rm in}=$100 GeV.} 
\end{table*}

\begin{table*}\label{table1}
  \begin{center}
    \begin{tabular}{|c|c|c|c|c|c|c|}\hline
      \hline 
       $x_{\rm e}$ (ionized   & Gas     &  Excitations     & Ionizations      &      Excitations  &    HE photons  & Energy from \\
      fraction)          & Heating & (Lyman-$\alpha$) & (H, He, HeII)    & ($E<10.2$ eV)   &   ($E> 10$ keV)   &   CMB\\
      \hline 
1.000e-04 &5.9519e-05$\pm$0e-00 &1.1637e-04$\pm$0e-00 &1.3531e-04$\pm$0e-00 &6.1362e-05$\pm$0e-00 &9.9961e-01$\pm$0e-00 &9.1146e-08$\pm$0e-00\\
2.779e-04 &6.8263e-05$\pm$0e-00 &1.1079e-04$\pm$0e-00 &1.3369e-04$\pm$0e-00 &5.8872e-05$\pm$0e-00 &9.9961e-01$\pm$0e-00 &9.1021e-08$\pm$0e-00\\
7.725e-04 &8.0141e-05$\pm$0e-00 &1.0461e-04$\pm$0e-00 &1.3102e-04$\pm$0e-00 &5.6055e-05$\pm$0e-00 &9.9961e-01$\pm$0e-00 &9.1010e-08$\pm$0e-00\\
2.147e-03 &9.6919e-05$\pm$0e-00 &9.7361e-05$\pm$0e-00 &1.2628e-04$\pm$0e-00 &5.2744e-05$\pm$0e-00 &9.9961e-01$\pm$0e-00 &9.1096e-08$\pm$0e-00\\
5.968e-03 &1.2010e-04$\pm$0e-00 &8.7357e-05$\pm$0e-00 &1.1660e-04$\pm$0e-00 &4.8193e-05$\pm$0e-00 &9.9961e-01$\pm$0e-00 &9.0909e-08$\pm$0e-00\\
1.658e-02 &1.5380e-04$\pm$0e-00 &7.4406e-05$\pm$0e-00 &1.0123e-04$\pm$0e-00 &4.2479e-05$\pm$0e-00 &9.9961e-01$\pm$0e-00 &9.1139e-08$\pm$0e-00\\
4.610e-02 &2.0188e-04$\pm$0e-00 &5.7279e-05$\pm$0e-00 &7.8283e-05$\pm$0e-00 &3.4933e-05$\pm$0e-00 &9.9961e-01$\pm$0e-00 &8.8266e-08$\pm$0e-00\\
1.281e-01 &2.6155e-04$\pm$0e-00 &3.6048e-05$\pm$0e-00 &4.8541e-05$\pm$0e-00 &2.5755e-05$\pm$0e-00 &9.9961e-01$\pm$0e-00 &9.1178e-08$\pm$0e-00\\
3.561e-01 &3.1947e-04$\pm$0e-00 &1.5889e-05$\pm$0e-00 &1.9965e-05$\pm$0e-00 &1.6965e-05$\pm$0e-00 &9.9961e-01$\pm$0e-00 &9.0696e-08$\pm$0e-00\\
9.900e-01 &3.5875e-04$\pm$0e-00 &2.0092e-06$\pm$0e-00 &1.4461e-07$\pm$0e-00 &1.0908e-05$\pm$0e-00 &9.9961e-01$\pm$0e-00 &9.0716e-08$\pm$0e-00\\
      \hline
    \end{tabular}
  \end{center}
  \caption{Same as table A1 but for energy $E_{\rm in}=$1 TeV.} 
\end{table*}

\begin{table*}\label{table1}
  \begin{center}
    \begin{tabular}{|c|c|c|c|c|c|c|}\hline
      \hline 
       $x_{\rm e}$ (ionized   & Gas     &  Excitations     & Ionizations      &      Excitations  &    HE photons  & Energy from \\
      fraction)          & Heating & (Lyman-$\alpha$) & (H, He, HeII)    & ($E<10.2$ eV)   &   ($E> 10$ keV)   &   CMB\\
      \hline 
1.000e-04&4.7949e-02$\pm$7e-04&1.0609e-01$\pm$1e-03&1.1792e-01$\pm$2e-03&6.6748e-01$\pm$3e-03&0.0000e-00$\pm$0e-00&6.0649e-02$\pm$3e-04\\
2.779e-04&5.5115e-02$\pm$1e-03&1.0305e-01$\pm$2e-03&1.1828e-01$\pm$2e-03&6.6338e-01$\pm$5e-03&0.0000e-00$\pm$0e-00&6.0253e-02$\pm$6e-04\\
7.725e-04&6.3611e-02$\pm$9e-04&9.7990e-02$\pm$2e-03&1.1581e-01$\pm$1e-03&6.6220e-01$\pm$3e-03&0.0000e-00$\pm$0e-00&6.0483e-02$\pm$3e-04\\
2.147e-03&7.4715e-02$\pm$2e-03&9.2606e-02$\pm$2e-03&1.1272e-01$\pm$2e-03&6.5958e-01$\pm$5e-03&0.0000e-00$\pm$0e-00&6.0460e-02$\pm$6e-04\\
5.968e-03&8.9678e-02$\pm$2e-03&8.5730e-02$\pm$2e-03&1.0748e-01$\pm$2e-03&6.5665e-01$\pm$5e-03&0.0000e-00$\pm$0e-00&6.0509e-02$\pm$6e-04\\
1.659e-02&1.0838e-01$\pm$7e-04&7.6995e-02$\pm$3e-04&9.9004e-02$\pm$7e-04&6.5495e-01$\pm$1e-03&0.0000e-00$\pm$0e-00&6.0697e-02$\pm$1e-04\\
4.611e-02&1.3406e-01$\pm$3e-03&6.8027e-02$\pm$1e-03&8.6996e-02$\pm$1e-03&6.5028e-01$\pm$2e-03&0.0000e-00$\pm$0e-00&6.0671e-02$\pm$2e-04\\
1.281e-01&1.7903e-01$\pm$1e-02&5.4899e-02$\pm$2e-03&7.0672e-02$\pm$2e-03&6.3574e-01$\pm$1e-02&0.0000e-00$\pm$0e-00&5.9656e-02$\pm$1e-03\\
3.562e-01&2.4571e-01$\pm$1e-02&3.0146e-02$\pm$2e-03&3.9021e-02$\pm$3e-03&6.2525e-01$\pm$1e-02&0.0000e-00$\pm$0e-00&5.9872e-02$\pm$1e-03\\
9.900e-01&3.2804e-01$\pm$2e-02&3.1720e-04$\pm$6e-05&3.7808e-04$\pm$5e-05&6.1190e-01$\pm$2e-02&0.0000e-00$\pm$0e-00&5.9360e-02$\pm$2e-03\\
      \hline
    \end{tabular}
  \end{center}
  \caption{Fraction of the energy $E_{\rm in}$ of a 1 MeV primary electron that is deposited into heat, ionizations, Ly$\alpha$ line radiation, 
photons with energy $E <$ 10.2 eV, photons with energies $E > 10$ keV due to Inverse Compton. The last column shows the total energy from the CMB 
photons before they were upscattered as a test of energy conservation. We consider here redshift $z=30$.} 
\end{table*}

\begin{table*}\label{table1}
  \begin{center}
    \begin{tabular}{|c|c|c|c|c|c|c|}\hline
      \hline 
       $x_{\rm e}$ (ionized   & Gas     &  Excitations     & Ionizations      &      Excitations  &    HE photons  & Energy from \\
      fraction)          & Heating & (Lyman-$\alpha$) & (H, He, HeII)    & ($E<10.2$ eV)   &   ($E> 10$ keV)   &   CMB\\
      \hline 
1.000e-04&6.5170e-02$\pm$2e-04&1.1944e-01$\pm$2e-04&2.6558e-02$\pm$2e-04&6.0778e-01$\pm$5e-04&0.0000e-00$\pm$0e-00&1.3215e-02$\pm$4e-05\\
2.779e-04&7.2613e-02$\pm$9e-05&1.1505e-01$\pm$2e-04&2.5641e-02$\pm$2e-04&6.0580e-01$\pm$4e-04&0.0000e-00$\pm$0e-00&1.3211e-02$\pm$2e-05\\
7.725e-04&8.2565e-02$\pm$2e-04&1.0961e-01$\pm$4e-04&2.3998e-02$\pm$2e-04&6.0305e-01$\pm$5e-04&0.0000e-00$\pm$0e-00&1.3209e-02$\pm$4e-05\\
2.147e-03&9.4603e-02$\pm$3e-04&1.0328e-01$\pm$3e-04&2.0942e-02$\pm$5e-04&6.0027e-01$\pm$9e-04&0.0000e-00$\pm$0e-00&1.3197e-02$\pm$8e-05\\
5.968e-03&1.0687e-01$\pm$2e-04&9.7519e-02$\pm$3e-04&1.7271e-02$\pm$2e-04&5.9751e-01$\pm$4e-04&0.0000e-00$\pm$0e-00&1.3194e-02$\pm$5e-05\\
1.659e-02&1.1754e-01$\pm$3e-04&9.2524e-02$\pm$2e-04&1.3634e-02$\pm$3e-04&5.9544e-01$\pm$7e-04&0.0000e-00$\pm$0e-00&1.3193e-02$\pm$8e-05\\
4.611e-02&1.2576e-01$\pm$5e-04&8.9044e-02$\pm$4e-04&1.0509e-02$\pm$3e-04&5.9370e-01$\pm$1e-03&0.0000e-00$\pm$0e-00&1.3200e-02$\pm$8e-05\\
1.281e-01&1.3280e-01$\pm$6e-04&8.6449e-02$\pm$4e-04&7.6575e-03$\pm$4e-04&5.9231e-01$\pm$1e-03&0.0000e-00$\pm$0e-00&1.3206e-02$\pm$8e-05\\
3.562e-01&1.4116e-01$\pm$2e-03&8.3597e-02$\pm$4e-04&4.2616e-03$\pm$4e-04&5.9028e-01$\pm$2e-03&0.0000e-00$\pm$0e-00&1.3106e-02$\pm$2e-04\\
9.900e-01&1.5355e-01$\pm$6e-03&8.0184e-02$\pm$3e-04&5.1136e-05$\pm$1e-05&5.8547e-01$\pm$6e-03&0.0000e-00$\pm$0e-00&1.3022e-02$\pm$2e-04\\
      \hline
    \end{tabular}
  \end{center}
  \caption{Same as table A8 but for energy $E_{\rm in}=$10 MeV.} 
\end{table*}

\clearpage

\begin{table*}\label{table1}
  \begin{center}
    \begin{tabular}{|c|c|c|c|c|c|c|}\hline
      \hline 
       $x_{\rm e}$ (ionized   & Gas     &  Excitations     & Ionizations      &      Excitations  &    HE photons  & Energy from \\
      fraction)          & Heating & (Lyman-$\alpha$) & (H, He, HeII)    & ($E<10.2$ eV)   &   ($E> 10$ keV)   &   CMB\\
      \hline 
1.000e-04&1.5026e-01$\pm$7e-05&2.7986e-01$\pm$1e-04&3.1049e-01$\pm$1e-04&1.8201e-01$\pm$8e-05&1.6601e-04$\pm$1e-04&1.4322e-03$\pm$0e-00\\
2.779e-04&1.7274e-01$\pm$7e-05&2.6661e-01$\pm$1e-04&3.0699e-01$\pm$9e-05&1.7611e-01$\pm$9e-05&2.9947e-04$\pm$1e-04&1.4327e-03$\pm$0e-00\\
7.725e-04&2.0442e-01$\pm$1e-04&2.5009e-01$\pm$2e-04&2.9964e-01$\pm$2e-04&1.6846e-01$\pm$5e-04&1.3948e-04$\pm$1e-04&1.4301e-03$\pm$0e-00\\
2.147e-03&2.4961e-01$\pm$1e-04&2.2910e-01$\pm$1e-04&2.8492e-01$\pm$1e-04&1.5900e-01$\pm$7e-05&2.0441e-04$\pm$1e-04&1.4348e-03$\pm$0e-00\\
5.968e-03&3.1527e-01$\pm$1e-04&2.0186e-01$\pm$1e-04&2.5877e-01$\pm$7e-05&1.4669e-01$\pm$8e-05&2.3742e-04$\pm$1e-04&1.4310e-03$\pm$0e-00\\
1.659e-02&4.0853e-01$\pm$1e-04&1.6631e-01$\pm$1e-04&2.1675e-01$\pm$1e-04&1.3091e-01$\pm$8e-05&1.9667e-04$\pm$1e-04&1.4290e-03$\pm$0e-00\\
4.611e-02&5.3057e-01$\pm$2e-04&1.2215e-01$\pm$1e-04&1.5794e-01$\pm$1e-04&1.1177e-01$\pm$7e-05&2.2250e-04$\pm$1e-04&1.4339e-03$\pm$0e-00\\
1.281e-01&6.6513e-01$\pm$2e-04&7.4742e-02$\pm$7e-05&9.1325e-02$\pm$1e-04&9.1234e-02$\pm$1e-04&2.0928e-04$\pm$2e-04&1.4275e-03$\pm$0e-00\\
3.562e-01&7.7729e-01$\pm$4e-04&3.5669e-02$\pm$9e-05&3.5254e-02$\pm$7e-05&7.4237e-02$\pm$4e-04&1.8391e-04$\pm$1e-04&1.4207e-03$\pm$2e-05\\
9.900e-01&8.4660e-01$\pm$2e-04&1.1518e-02$\pm$3e-05&2.4533e-04$\pm$0e-00&6.3929e-02$\pm$3e-04&3.6417e-04$\pm$2e-04&1.4242e-03$\pm$3e-05\\
      \hline
    \end{tabular}
  \end{center}
  \caption{Same as table A8 but for energy $E_{\rm in}=$100 MeV.} 
\end{table*}

\begin{table*}\label{table1}
  \begin{center}
    \begin{tabular}{|c|c|c|c|c|c|c|}\hline
      \hline 
       $x_{\rm e}$ (ionized   & Gas     &  Excitations     & Ionizations      &      Excitations  &    HE photons  & Energy from \\
      fraction)          & Heating & (Lyman-$\alpha$) & (H, He, HeII)    & ($E<10.2$ eV)   &   ($E> 10$ keV)   &   CMB\\
      \hline 
1.000e-04&3.4982e-02$\pm$1e-04&6.9174e-02$\pm$3e-04&8.0259e-02$\pm$3e-04&3.6397e-02$\pm$1e-04&7.7092e-01$\pm$9e-04&1.4423e-04$\pm$0e-00\\
2.779e-04&4.0132e-02$\pm$1e-04&6.6110e-02$\pm$2e-04&7.9542e-02$\pm$3e-04&3.5038e-02$\pm$1e-04&7.7090e-01$\pm$8e-04&1.4438e-04$\pm$0e-00\\
7.725e-04&4.6999e-02$\pm$2e-04&6.2419e-02$\pm$3e-04&7.7974e-02$\pm$4e-04&3.3367e-02$\pm$1e-04&7.7096e-01$\pm$1e-03&1.4452e-04$\pm$0e-00\\
2.147e-03&5.6658e-02$\pm$1e-04&5.8146e-02$\pm$2e-04&7.5226e-02$\pm$2e-04&3.1410e-02$\pm$7e-05&7.7030e-01$\pm$6e-04&1.4378e-04$\pm$0e-00\\
5.968e-03&6.9999e-02$\pm$2e-04&5.2293e-02$\pm$2e-04&6.9671e-02$\pm$2e-04&2.8754e-02$\pm$8e-05&7.7101e-01$\pm$6e-04&1.4367e-04$\pm$0e-00\\
1.659e-02&8.9393e-02$\pm$2e-04&4.4778e-02$\pm$1e-04&6.0818e-02$\pm$1e-04&2.5424e-02$\pm$4e-05&7.7129e-01$\pm$5e-04&1.4431e-04$\pm$0e-00\\
4.611e-02&1.1757e-01$\pm$4e-04&3.4888e-02$\pm$1e-04&4.7676e-02$\pm$2e-04&2.1098e-02$\pm$6e-05&7.7046e-01$\pm$8e-04&1.4457e-04$\pm$0e-00\\
1.281e-01&1.5299e-01$\pm$3e-04&2.2289e-02$\pm$7e-05&3.0077e-02$\pm$9e-05&1.5611e-02$\pm$3e-05&7.7071e-01$\pm$5e-04&1.4451e-04$\pm$0e-00\\
3.562e-01&1.8792e-01$\pm$4e-04&9.9416e-03$\pm$4e-05&1.2576e-02$\pm$4e-05&1.0227e-02$\pm$3e-05&7.7102e-01$\pm$5e-04&1.4431e-04$\pm$0e-00\\
9.900e-01&2.1305e-01$\pm$6e-04&1.1995e-03$\pm$0e-00&9.3508e-05$\pm$0e-00&6.3982e-03$\pm$5e-05&7.7095e-01$\pm$6e-04&1.4076e-04$\pm$0e-00\\
    \hline
    \end{tabular}
  \end{center}
  \caption{Same as table A8 but for energy $E_{\rm in}=$1 GeV.} 
\end{table*}

\begin{table*}\label{table1}
  \begin{center}
    \begin{tabular}{|c|c|c|c|c|c|c|}\hline
      \hline 
       $x_{\rm e}$ (ionized   & Gas     &  Excitations     & Ionizations      &      Excitations  &    HE photons  & Energy from \\
      fraction)          & Heating & (Lyman-$\alpha$) & (H, He, HeII)    & ($E<10.2$ eV)   &   ($E> 10$ keV)   &   CMB\\
      \hline 
1.000e-04&3.5042e-03$\pm$1e-05&6.9298e-03$\pm$3e-05&8.0414e-03$\pm$3e-05&3.6454e-03$\pm$1e-05&9.7705e-01$\pm$9e-05&1.4436e-05$\pm$0e-00\\
2.779e-04&4.0274e-03$\pm$1e-05&6.6343e-03$\pm$2e-05&7.9838e-03$\pm$3e-05&3.5137e-03$\pm$0e-00&9.7701e-01$\pm$7e-05&1.4419e-05$\pm$0e-00\\
7.725e-04&4.7147e-03$\pm$0e-00&6.2647e-03$\pm$1e-05&7.8255e-03$\pm$2e-05&3.3466e-03$\pm$0e-00&9.7702e-01$\pm$5e-05&1.4435e-05$\pm$0e-00\\
2.147e-03&5.6701e-03$\pm$1e-05&5.8218e-03$\pm$2e-05&7.5302e-03$\pm$2e-05&3.1443e-03$\pm$0e-00&9.7701e-01$\pm$6e-05&1.4418e-05$\pm$0e-00\\
5.968e-03&7.0342e-03$\pm$2e-05&5.2585e-03$\pm$2e-05&7.0073e-03$\pm$2e-05&2.8888e-03$\pm$0e-00&9.7698e-01$\pm$6e-05&1.4414e-05$\pm$0e-00\\
1.659e-02&8.9826e-03$\pm$3e-05&4.5036e-03$\pm$2e-05&6.1184e-03$\pm$3e-05&2.5537e-03$\pm$0e-00&9.7701e-01$\pm$9e-05&1.4426e-05$\pm$0e-00\\
4.611e-02&1.1768e-02$\pm$2e-05&3.4923e-03$\pm$0e-00&4.7737e-03$\pm$1e-05&2.1111e-03$\pm$0e-00&9.7702e-01$\pm$5e-05&1.4440e-05$\pm$0e-00\\
1.281e-01&1.5347e-02$\pm$5e-05&2.2383e-03$\pm$0e-00&3.0212e-03$\pm$1e-05&1.5651e-03$\pm$0e-00&9.7700e-01$\pm$7e-05&1.4447e-05$\pm$0e-00\\
3.562e-01&1.8853e-02$\pm$7e-05&9.9756e-04$\pm$0e-00&1.2631e-03$\pm$0e-00&1.0237e-03$\pm$0e-00&9.7703e-01$\pm$9e-05&1.4331e-05$\pm$0e-00\\
9.900e-01&2.1387e-02$\pm$4e-05&1.1971e-04$\pm$0e-00&9.3111e-06$\pm$0e-00&6.3960e-04$\pm$0e-00&9.7701e-01$\pm$4e-05&1.4214e-05$\pm$0e-00\\
     \hline
    \end{tabular}
  \end{center}
  \caption{Same as table A8 but for energy $E_{\rm in}=$10 GeV.} 
\end{table*}

\begin{table*}\label{table1}
  \begin{center}
    \begin{tabular}{|c|c|c|c|c|c|c|}\hline
      \hline 
       $x_{\rm e}$ (ionized   & Gas     &  Excitations     & Ionizations      &      Excitations  &    HE photons  & Energy from \\
      fraction)          & Heating & (Lyman-$\alpha$) & (H, He, HeII)    & ($E<10.2$ eV)   &   ($E> 10$ keV)   &   CMB\\
      \hline 
1.000e-04&3.5090e-04$\pm$0e-00&6.9417e-04$\pm$0e-00&8.0536e-04$\pm$0e-00&3.6504e-04$\pm$0e-00&9.9770e-01$\pm$1e-05&1.4428e-06$\pm$0e-00\\
2.779e-04&4.0318e-04$\pm$0e-00&6.6436e-04$\pm$0e-00&7.9940e-04$\pm$0e-00&3.5179e-04$\pm$0e-00&9.9770e-01$\pm$0e-00&1.4459e-06$\pm$0e-00\\
7.725e-04&4.7170e-04$\pm$0e-00&6.2667e-04$\pm$0e-00&7.8295e-04$\pm$0e-00&3.3443e-04$\pm$0e-00&9.9770e-01$\pm$0e-00&1.4504e-06$\pm$0e-00\\
2.147e-03&5.6732e-04$\pm$0e-00&5.8250e-04$\pm$0e-00&7.5359e-04$\pm$0e-00&3.1450e-04$\pm$0e-00&9.9770e-01$\pm$0e-00&1.4379e-06$\pm$0e-00\\
5.968e-03&7.0301e-04$\pm$0e-00&5.2560e-04$\pm$0e-00&7.0031e-04$\pm$0e-00&2.8848e-04$\pm$0e-00&9.9770e-01$\pm$0e-00&1.4240e-06$\pm$0e-00\\
1.659e-02&8.9763e-04$\pm$0e-00&4.4973e-04$\pm$0e-00&6.1114e-04$\pm$0e-00&2.5509e-04$\pm$0e-00&9.9770e-01$\pm$0e-00&1.4463e-06$\pm$0e-00\\
4.611e-02&1.1773e-03$\pm$0e-00&3.4952e-04$\pm$0e-00&4.7774e-04$\pm$0e-00&2.1095e-04$\pm$0e-00&9.9770e-01$\pm$0e-00&1.4110e-06$\pm$0e-00\\
1.281e-01&1.5355e-03$\pm$0e-00&2.2404e-04$\pm$0e-00&3.0231e-04$\pm$0e-00&1.5653e-04$\pm$0e-00&9.9770e-01$\pm$0e-00&1.4356e-06$\pm$0e-00\\
3.562e-01&1.8874e-03$\pm$0e-00&9.9928e-05$\pm$0e-00&1.2646e-04$\pm$0e-00&1.0260e-04$\pm$0e-00&9.9770e-01$\pm$0e-00&1.4491e-06$\pm$0e-00\\
9.900e-01&2.1359e-03$\pm$0e-00&1.2005e-05$\pm$0e-00&9.3136e-07$\pm$0e-00&6.4095e-05$\pm$0e-00&9.9770e-01$\pm$0e-00&1.4297e-06$\pm$0e-00\\
    \hline
    \end{tabular}
  \end{center}
  \caption{Same as table A8 but for energy $E_{\rm in}=$100 GeV.} 
\end{table*}

\begin{table*}\label{table1}
  \begin{center}
    \begin{tabular}{|c|c|c|c|c|c|c|}\hline
      \hline 
       $x_{\rm e}$ (ionized   & Gas     &  Excitations     & Ionizations      &      Excitations  &    HE photons  & Energy from \\
      fraction)          & Heating & (Lyman-$\alpha$) & (H, He, HeII)    & ($E<10.2$ eV)   &   ($E> 10$ keV)   &   CMB\\
      \hline 
1.000e-04&3.5062e-05$\pm$0e-00&6.9376e-05$\pm$0e-00&8.0463e-05$\pm$0e-00&3.6491e-05$\pm$0e-00&9.9977e-01$\pm$0e-00&1.4450e-07$\pm$0e-00\\
2.779e-04&4.0301e-05$\pm$0e-00&6.6415e-05$\pm$0e-00&7.9895e-05$\pm$0e-00&3.5172e-05$\pm$0e-00&9.9977e-01$\pm$0e-00&1.4458e-07$\pm$0e-00\\
7.725e-04&4.7208e-05$\pm$0e-00&6.2730e-05$\pm$0e-00&7.8365e-05$\pm$0e-00&3.3504e-05$\pm$0e-00&9.9977e-01$\pm$0e-00&1.4457e-07$\pm$0e-00\\
2.147e-03&5.6731e-05$\pm$0e-00&5.8230e-05$\pm$0e-00&7.5355e-05$\pm$0e-00&3.1444e-05$\pm$0e-00&9.9977e-01$\pm$0e-00&1.4372e-07$\pm$0e-00\\
5.968e-03&7.0191e-05$\pm$0e-00&5.2465e-05$\pm$0e-00&6.9898e-05$\pm$0e-00&2.8839e-05$\pm$0e-00&9.9977e-01$\pm$0e-00&1.4445e-07$\pm$0e-00\\
1.659e-02&8.9687e-05$\pm$0e-00&4.4938e-05$\pm$0e-00&6.1055e-05$\pm$0e-00&2.5490e-05$\pm$0e-00&9.9977e-01$\pm$0e-00&1.4460e-07$\pm$0e-00\\
4.611e-02&1.1769e-04$\pm$0e-00&3.4929e-05$\pm$0e-00&4.7759e-05$\pm$0e-00&2.1092e-05$\pm$0e-00&9.9977e-01$\pm$0e-00&1.4348e-07$\pm$0e-00\\
1.281e-01&1.5332e-04$\pm$0e-00&2.2360e-05$\pm$0e-00&3.0174e-05$\pm$0e-00&1.5647e-05$\pm$0e-00&9.9977e-01$\pm$0e-00&1.4463e-07$\pm$0e-00\\
3.562e-01&1.8862e-04$\pm$0e-00&9.9770e-06$\pm$0e-00&1.2636e-05$\pm$0e-00&1.0248e-05$\pm$0e-00&9.9977e-01$\pm$0e-00&1.4552e-07$\pm$0e-00\\
9.900e-01&2.1377e-04$\pm$0e-00&1.1994e-06$\pm$0e-00&9.3606e-08$\pm$0e-00&6.4119e-06$\pm$0e-00&9.9977e-01$\pm$0e-00&1.4344e-07$\pm$0e-00\\
    \hline
    \end{tabular}
  \end{center}
  \caption{Same as table A8 but for energy $E_{\rm in}=$1 TeV.} 
\end{table*}

\begin{table*}\label{table1}
  \begin{center}
    \begin{tabular}{|c|c|c|c|c|c|c|}\hline
      \hline 
       $x_{\rm e}$ (ionized   & Gas     &  Excitations     & Ionizations      &      Excitations  &    HE photons  & Energy from \\
      fraction)          & Heating & (Lyman-$\alpha$) & (H, He, HeII)    & ($E<10.2$ eV)   &   ($E> 10$ keV)   &   CMB\\
      \hline 
1.000e-04&2.1464e-02$\pm$6e-04&4.7691e-02$\pm$1e-03&5.3596e-02$\pm$2e-03&7.8527e-01$\pm$3e-03&0.0000e-00$\pm$0e-00&9.2015e-02$\pm$6e-04\\
2.779e-04&2.4840e-02$\pm$2e-03&4.6612e-02$\pm$4e-03&5.3886e-02$\pm$5e-03&7.8269e-01$\pm$1e-02&0.0000e-00$\pm$0e-00&9.1998e-02$\pm$1e-03\\
7.725e-04&2.7823e-02$\pm$4e-04&4.2640e-02$\pm$7e-04&5.1324e-02$\pm$6e-04&7.8573e-01$\pm$1e-03&0.0000e-00$\pm$0e-00&9.2518e-02$\pm$2e-04\\
2.147e-03&3.3214e-02$\pm$7e-04&4.0728e-02$\pm$1e-03&5.0252e-02$\pm$8e-04&7.8351e-01$\pm$2e-03&0.0000e-00$\pm$0e-00&9.2312e-02$\pm$3e-04\\
5.968e-03&3.9490e-02$\pm$1e-03&3.7864e-02$\pm$1e-03&4.7959e-02$\pm$1e-03&7.8240e-01$\pm$2e-03&0.0000e-00$\pm$0e-00&9.2316e-02$\pm$4e-04\\
1.659e-02&4.8279e-02$\pm$2e-03&3.4619e-02$\pm$1e-03&4.4487e-02$\pm$2e-03&7.8042e-01$\pm$5e-03&0.0000e-00$\pm$0e-00&9.2213e-02$\pm$9e-04\\
4.611e-02&6.0673e-02$\pm$3e-03&3.0751e-02$\pm$2e-03&4.0295e-02$\pm$2e-03&7.7664e-01$\pm$6e-03&0.0000e-00$\pm$0e-00&9.1642e-02$\pm$2e-03\\
1.281e-01&8.0938e-02$\pm$4e-03&2.3972e-02$\pm$2e-03&3.1881e-02$\pm$2e-03&7.7194e-01$\pm$6e-03&0.0000e-00$\pm$0e-00&9.1268e-02$\pm$1e-03\\
3.562e-01&1.0719e-01$\pm$7e-03&1.3114e-02$\pm$7e-04&1.7022e-02$\pm$9e-04&7.7065e-01$\pm$7e-03&0.0000e-00$\pm$0e-00&9.2023e-02$\pm$1e-03\\
9.900e-01&1.7492e-01$\pm$3e-02&1.4520e-04$\pm$5e-05&1.9040e-04$\pm$6e-05&7.3680e-01$\pm$3e-02&0.0000e-00$\pm$0e-00&8.7949e-02$\pm$4e-03\\
     \hline
    \end{tabular}
  \end{center}
  \caption{Fraction of the energy $E_{\rm in}$ of a 1 MeV primary electron that is deposited into heat, ionizations, Ly$\alpha$ line radiation, 
photons with energy $E <$ 10.2 eV, photons with energies $E > 10$ keV due to Inverse Compton. The last column shows the total energy from the CMB 
photons before they were upscattered as a test of energy conservation. We consider here redshift $z=50$.} 
\end{table*}

\clearpage

\begin{table*}\label{table1}
  \begin{center}
    \begin{tabular}{|c|c|c|c|c|c|c|}\hline
      \hline 
       $x_{\rm e}$ (ionized   & Gas     &  Excitations     & Ionizations      &      Excitations  &    HE photons  & Energy from \\
      fraction)          & Heating & (Lyman-$\alpha$) & (H, He, HeII)    & ($E<10.2$ eV)   &   ($E> 10$ keV)   &   CMB\\
      \hline 
1.000e-04&8.7513e-02$\pm$2e-04&1.3675e-01$\pm$4e-04&4.1413e-02$\pm$3e-04&5.0827e-01$\pm$7e-04&0.0000e-00$\pm$0e-00&1.6381e-02$\pm$8e-05\\
2.779e-04&9.8787e-02$\pm$2e-04&1.3023e-01$\pm$5e-04&3.9967e-02$\pm$5e-04&5.0524e-01$\pm$9e-04&0.0000e-00$\pm$0e-00&1.6292e-02$\pm$2e-04\\
7.725e-04&1.1476e-01$\pm$2e-04&1.2139e-01$\pm$3e-04&3.6426e-02$\pm$2e-04&5.0138e-01$\pm$6e-04&0.0000e-00$\pm$0e-00&1.6366e-02$\pm$7e-05\\
2.147e-03&1.3547e-01$\pm$4e-04&1.1109e-01$\pm$5e-04&3.1073e-02$\pm$3e-04&4.9636e-01$\pm$1e-03&0.0000e-00$\pm$0e-00&1.6356e-02$\pm$1e-04\\
5.968e-03&1.5874e-01$\pm$4e-04&1.0003e-01$\pm$3e-04&2.3829e-02$\pm$3e-04&4.9143e-01$\pm$7e-04&0.0000e-00$\pm$0e-00&1.6358e-02$\pm$1e-04\\
1.659e-02&1.8046e-01$\pm$5e-04&9.0628e-02$\pm$4e-04&1.5941e-02$\pm$4e-04&4.8689e-01$\pm$1e-03&0.0000e-00$\pm$0e-00&1.6332e-02$\pm$2e-04\\
4.611e-02&1.9638e-01$\pm$5e-04&8.3771e-02$\pm$3e-04&9.5488e-03$\pm$3e-04&4.8417e-01$\pm$8e-04&0.0000e-00$\pm$0e-00&1.6391e-02$\pm$9e-05\\
1.281e-01&2.0678e-01$\pm$5e-04&7.9688e-02$\pm$3e-04&5.3475e-03$\pm$2e-04&4.8205e-01$\pm$6e-04&0.0000e-00$\pm$0e-00&1.6358e-02$\pm$8e-05\\
3.562e-01&2.1389e-01$\pm$9e-04&7.7199e-02$\pm$3e-04&2.4317e-03$\pm$1e-04&4.8065e-01$\pm$1e-03&0.0000e-00$\pm$0e-00&1.6262e-02$\pm$2e-04\\
9.900e-01&2.2143e-01$\pm$3e-03&7.5084e-02$\pm$3e-04&2.1216e-05$\pm$0e-00&4.7777e-01$\pm$2e-03&0.0000e-00$\pm$0e-00&1.5950e-02$\pm$4e-04\\
     \hline
    \end{tabular}
  \end{center}
  \caption{Same as table A15 but for energy $E_{\rm in}=$10 MeV.} 
\end{table*}

\begin{table*}\label{table1}
  \begin{center}
    \begin{tabular}{|c|c|c|c|c|c|c|}\hline
      \hline 
       $x_{\rm e}$ (ionized   & Gas     &  Excitations     & Ionizations      &      Excitations  &    HE photons  & Energy from \\
      fraction)          & Heating & (Lyman-$\alpha$) & (H, He, HeII)    & ($E<10.2$ eV)   &   ($E> 10$ keV)   &   CMB\\
      \hline 
1.000e-04&1.5014e-01$\pm$5e-05&2.8695e-01$\pm$2e-04&3.2422e-01$\pm$1e-04&1.7231e-01$\pm$1e-04&3.3436e-03$\pm$5e-04&1.7491e-03$\pm$0e-00\\
2.779e-04&1.7239e-01$\pm$1e-04&2.7407e-01$\pm$2e-04&3.2116e-01$\pm$3e-04&1.6658e-01$\pm$1e-04&2.7406e-03$\pm$7e-04&1.7468e-03$\pm$0e-00\\
7.725e-04&2.0298e-01$\pm$2e-04&2.5781e-01$\pm$2e-04&3.1410e-01$\pm$3e-04&1.5916e-01$\pm$2e-04&2.8659e-03$\pm$7e-04&1.7452e-03$\pm$1e-05\\
2.147e-03&2.4624e-01$\pm$2e-04&2.3739e-01$\pm$2e-04&3.0016e-01$\pm$2e-04&1.4987e-01$\pm$8e-05&3.3213e-03$\pm$5e-04&1.7494e-03$\pm$0e-00\\
5.968e-03&3.0900e-01$\pm$2e-04&2.1138e-01$\pm$2e-04&2.7551e-01$\pm$2e-04&1.3816e-01$\pm$1e-04&2.9106e-03$\pm$6e-04&1.7525e-03$\pm$0e-00\\
1.659e-02&3.9881e-01$\pm$3e-04&1.7692e-01$\pm$1e-04&2.3497e-01$\pm$2e-04&1.2279e-01$\pm$1e-04&3.3355e-03$\pm$6e-04&1.7415e-03$\pm$2e-05\\
4.611e-02&5.2105e-01$\pm$4e-04&1.3283e-01$\pm$2e-04&1.7634e-01$\pm$2e-04&1.0361e-01$\pm$1e-04&2.9545e-03$\pm$5e-04&1.7451e-03$\pm$1e-05\\
1.281e-01&6.6337e-01$\pm$4e-04&8.2768e-02$\pm$1e-04&1.0590e-01$\pm$1e-04&8.1943e-02$\pm$1e-04&2.7966e-03$\pm$5e-04&1.7449e-03$\pm$1e-05\\
3.562e-01&7.9064e-01$\pm$3e-04&3.8281e-02$\pm$4e-05&4.2318e-02$\pm$8e-05&6.2668e-02$\pm$7e-05&2.9297e-03$\pm$3e-04&1.7313e-03$\pm$2e-05\\
9.900e-01&8.7471e-01$\pm$7e-04&9.1588e-03$\pm$3e-05&3.0206e-04$\pm$0e-00&4.9806e-02$\pm$5e-04&2.8655e-03$\pm$5e-04&1.7198e-03$\pm$3e-05\\
 \hline
    \end{tabular}
  \end{center}
  \caption{Same as table A15 but for energy $E_{\rm in}=$100 MeV.} 
\end{table*}

\begin{table*}\label{table2}
  \begin{center}
    \begin{tabular}{|c|c|c|c|c|c|c|}\hline
      \hline 
       $x_{\rm e}$ (ionized   & Gas     &  Excitations     & Ionizations      &      Excitations  &    HE photons  & Energy from \\
      fraction)          & Heating & (Lyman-$\alpha$) & (H, He, HeII)    & ($E<10.2$ eV)   &   ($E> 10$ keV)   &   CMB\\
      \hline 
1.000e-04&2.7571e-02$\pm$1e-04&5.4869e-02$\pm$2e-04&6.3583e-02$\pm$2e-04&2.8802e-02$\pm$8e-05&8.1858e-01$\pm$6e-04&1.7572e-04$\pm$0e-00\\
2.779e-04&3.1566e-02$\pm$9e-05&5.2382e-02$\pm$2e-04&6.2950e-02$\pm$2e-04&2.7704e-02$\pm$8e-05&8.1880e-01$\pm$5e-04&1.7598e-04$\pm$0e-00\\
7.725e-04&3.7050e-02$\pm$2e-04&4.9653e-02$\pm$3e-04&6.1947e-02$\pm$4e-04&2.6463e-02$\pm$1e-04&8.1828e-01$\pm$1e-03&1.7542e-04$\pm$0e-00\\
2.147e-03&4.4321e-02$\pm$1e-04&4.5970e-02$\pm$1e-04&5.9412e-02$\pm$2e-04&2.4788e-02$\pm$6e-05&8.1892e-01$\pm$5e-04&1.7634e-04$\pm$0e-00\\
5.968e-03&5.5067e-02$\pm$6e-05&4.1726e-02$\pm$7e-05&5.5554e-02$\pm$9e-05&2.2863e-02$\pm$3e-05&8.1818e-01$\pm$2e-04&1.7588e-04$\pm$0e-00\\
1.659e-02&7.0094e-02$\pm$2e-04&3.5732e-02$\pm$1e-04&4.8510e-02$\pm$2e-04&2.0198e-02$\pm$6e-05&8.1885e-01$\pm$6e-04&1.7644e-04$\pm$0e-00\\
4.611e-02&9.2022e-02$\pm$3e-04&2.7938e-02$\pm$1e-04&3.8190e-02$\pm$2e-04&1.6777e-02$\pm$5e-05&8.1844e-01$\pm$7e-04&1.7566e-04$\pm$0e-00\\
1.281e-01&1.1988e-01$\pm$5e-04&1.7980e-02$\pm$1e-04&2.4284e-02$\pm$1e-04&1.2442e-02$\pm$4e-05&8.1878e-01$\pm$8e-04&1.7535e-04$\pm$0e-00\\
3.562e-01&1.4856e-01$\pm$5e-04&8.1154e-03$\pm$3e-05&1.0324e-02$\pm$4e-05&8.1512e-03$\pm$2e-05&8.1822e-01$\pm$5e-04&1.7523e-04$\pm$0e-00\\
9.900e-01&1.6831e-01$\pm$7e-04&9.4852e-04$\pm$0e-00&7.6582e-05$\pm$0e-00&5.0211e-03$\pm$1e-05&8.1901e-01$\pm$7e-04&1.7432e-04$\pm$0e-00\\
\hline
    \end{tabular}
  \end{center}
  \caption{Same as table A15 but for energy $E_{\rm in}=$1 GeV.} 
\end{table*}

\begin{table*}\label{table1}
  \begin{center}
    \begin{tabular}{|c|c|c|c|c|c|c|}\hline
      \hline 
       $x_{\rm e}$ (ionized   & Gas     &  Excitations     & Ionizations      &      Excitations  &    HE photons  & Energy from \\
      fraction)          & Heating & (Lyman-$\alpha$) & (H, He, HeII)    & ($E<10.2$ eV)   &   ($E> 10$ keV)   &   CMB\\
      \hline 
1.000e-04&2.7587e-03$\pm$0e-00&5.4925e-03$\pm$2e-05&6.3624e-03$\pm$2e-05&2.8830e-03$\pm$0e-00&9.8184e-01$\pm$5e-05&1.7624e-05$\pm$0e-00\\
2.779e-04&3.1693e-03$\pm$1e-05&5.2615e-03$\pm$2e-05&6.3222e-03$\pm$3e-05&2.7809e-03$\pm$0e-00&9.8181e-01$\pm$7e-05&1.7635e-05$\pm$0e-00\\
7.725e-04&3.7000e-03$\pm$2e-05&4.9611e-03$\pm$2e-05&6.1873e-03$\pm$3e-05&2.6446e-03$\pm$1e-05&9.8185e-01$\pm$8e-05&1.7632e-05$\pm$0e-00\\
2.147e-03&4.4451e-03$\pm$1e-05&4.6130e-03$\pm$2e-05&5.9606e-03$\pm$2e-05&2.4861e-03$\pm$0e-00&9.8184e-01$\pm$5e-05&1.7635e-05$\pm$0e-00\\
5.968e-03&5.5048e-03$\pm$2e-05&4.1708e-03$\pm$2e-05&5.5525e-03$\pm$3e-05&2.2847e-03$\pm$0e-00&9.8183e-01$\pm$7e-05&1.7543e-05$\pm$0e-00\\
1.659e-02&7.0190e-03$\pm$3e-05&3.5792e-03$\pm$1e-05&4.8620e-03$\pm$2e-05&2.0218e-03$\pm$0e-00&9.8186e-01$\pm$7e-05&1.7487e-05$\pm$0e-00\\
4.611e-02&9.2132e-03$\pm$2e-05&2.7984e-03$\pm$0e-00&3.8265e-03$\pm$0e-00&1.6799e-03$\pm$0e-00&9.8182e-01$\pm$3e-05&1.7578e-05$\pm$0e-00\\
1.281e-01&1.1999e-02$\pm$3e-05&1.8000e-03$\pm$0e-00&2.4321e-03$\pm$0e-00&1.2448e-03$\pm$0e-00&9.8186e-01$\pm$5e-05&1.7527e-05$\pm$0e-00\\
3.562e-01&1.4824e-02$\pm$5e-05&8.1001e-04$\pm$0e-00&1.0296e-03$\pm$0e-00&8.1436e-04$\pm$0e-00&9.8186e-01$\pm$6e-05&1.7550e-05$\pm$0e-00\\
9.900e-01&1.6892e-02$\pm$5e-05&9.5038e-05$\pm$0e-00&7.6904e-06$\pm$0e-00&5.0075e-04$\pm$0e-00&9.8184e-01$\pm$5e-05&1.7205e-05$\pm$0e-00\\
\hline
    \end{tabular}
  \end{center}
  \caption{Same as table A15 but for energy $E_{\rm in}=$10 GeV.} 
\end{table*}

\begin{table*}\label{table1}
  \begin{center}
    \begin{tabular}{|c|c|c|c|c|c|c|}\hline
      \hline 
       $x_{\rm e}$ (ionized   & Gas     &  Excitations     & Ionizations      &      Excitations  &    HE photons  & Energy from \\
      fraction)          & Heating & (Lyman-$\alpha$) & (H, He, HeII)    & ($E<10.2$ eV)   &   ($E> 10$ keV)   &   CMB\\
      \hline 
1.000e-04&2.7558e-04$\pm$0e-00&5.4850e-04$\pm$0e-00&6.3558e-04$\pm$0e-00&2.8793e-04$\pm$0e-00&9.9819e-01$\pm$0e-00&1.7582e-06$\pm$0e-00\\
2.779e-04&3.1645e-04$\pm$0e-00&5.2523e-04$\pm$0e-00&6.3109e-04$\pm$0e-00&2.7758e-04$\pm$0e-00&9.9818e-01$\pm$0e-00&1.7497e-06$\pm$0e-00\\
7.725e-04&3.7029e-04$\pm$0e-00&4.9635e-04$\pm$0e-00&6.1930e-04$\pm$0e-00&2.6459e-04$\pm$0e-00&9.9818e-01$\pm$0e-00&1.7645e-06$\pm$0e-00\\
2.147e-03&4.4394e-04$\pm$0e-00&4.6070e-04$\pm$0e-00&5.9514e-04$\pm$0e-00&2.4820e-04$\pm$0e-00&9.9819e-01$\pm$0e-00&1.7453e-06$\pm$0e-00\\
5.968e-03&5.5019e-04$\pm$0e-00&4.1694e-04$\pm$0e-00&5.5505e-04$\pm$0e-00&2.2838e-04$\pm$0e-00&9.9818e-01$\pm$0e-00&1.7511e-06$\pm$0e-00\\
1.659e-02&7.0310e-04$\pm$0e-00&3.5880e-04$\pm$0e-00&4.8717e-04$\pm$0e-00&2.0263e-04$\pm$0e-00&9.9818e-01$\pm$0e-00&1.7601e-06$\pm$0e-00\\
4.611e-02&9.2039e-04$\pm$0e-00&2.7953e-04$\pm$0e-00&3.8212e-04$\pm$0e-00&1.6789e-04$\pm$0e-00&9.9818e-01$\pm$0e-00&1.7616e-06$\pm$0e-00\\
1.281e-01&1.2018e-03$\pm$0e-00&1.8043e-04$\pm$0e-00&2.4363e-04$\pm$0e-00&1.2460e-04$\pm$0e-00&9.9818e-01$\pm$0e-00&1.7558e-06$\pm$0e-00\\
3.562e-01&1.4845e-03$\pm$0e-00&8.1159e-05$\pm$0e-00&1.0314e-04$\pm$0e-00&8.1416e-05$\pm$0e-00&9.9818e-01$\pm$0e-00&1.7420e-06$\pm$0e-00\\
9.900e-01&1.6892e-03$\pm$0e-00&9.4977e-06$\pm$0e-00&7.7807e-07$\pm$0e-00&5.0153e-05$\pm$0e-00&9.9818e-01$\pm$0e-00&1.7333e-06$\pm$0e-00\\
\hline
    \end{tabular}
  \end{center}
  \caption{Same as table A15 but for energy $E_{\rm in}=$100 GeV.} 
\end{table*}

\begin{table*}\label{table1}
  \begin{center}
    \begin{tabular}{|c|c|c|c|c|c|c|}\hline
      \hline 
       $x_{\rm e}$ (ionized   & Gas     &  Excitations     & Ionizations      &      Excitations  &    HE photons  & Energy from \\
      fraction)          & Heating & (Lyman-$\alpha$) & (H, He, HeII)    & ($E<10.2$ eV)   &   ($E> 10$ keV)   &   CMB\\
      \hline 
1.000e-04&2.7539e-05$\pm$0e-00&5.4811e-05$\pm$0e-00&6.3515e-05$\pm$0e-00&2.8783e-05$\pm$0e-00&9.9982e-01$\pm$0e-00&1.7656e-07$\pm$0e-00\\
2.779e-04&3.1624e-05$\pm$0e-00&5.2490e-05$\pm$0e-00&6.3062e-05$\pm$0e-00&2.7749e-05$\pm$0e-00&9.9982e-01$\pm$0e-00&1.7624e-07$\pm$0e-00\\
7.725e-04&3.7066e-05$\pm$0e-00&4.9683e-05$\pm$0e-00&6.1980e-05$\pm$0e-00&2.6472e-05$\pm$0e-00&9.9982e-01$\pm$0e-00&1.7556e-07$\pm$0e-00\\
2.147e-03&4.4352e-05$\pm$0e-00&4.6031e-05$\pm$0e-00&5.9463e-05$\pm$0e-00&2.4812e-05$\pm$0e-00&9.9982e-01$\pm$0e-00&1.7623e-07$\pm$0e-00\\
5.968e-03&5.4933e-05$\pm$0e-00&4.1606e-05$\pm$0e-00&5.5403e-05$\pm$0e-00&2.2807e-05$\pm$0e-00&9.9982e-01$\pm$0e-00&1.7616e-07$\pm$0e-00\\
1.659e-02&7.0164e-05$\pm$0e-00&3.5759e-05$\pm$0e-00&4.8579e-05$\pm$0e-00&2.0173e-05$\pm$0e-00&9.9982e-01$\pm$0e-00&1.7780e-07$\pm$0e-00\\
4.611e-02&9.2019e-05$\pm$0e-00&2.7930e-05$\pm$0e-00&3.8186e-05$\pm$0e-00&1.6780e-05$\pm$0e-00&9.9982e-01$\pm$0e-00&1.7657e-07$\pm$0e-00\\
1.281e-01&1.2000e-04$\pm$0e-00&1.7997e-05$\pm$0e-00&2.4338e-05$\pm$0e-00&1.2456e-05$\pm$0e-00&9.9982e-01$\pm$0e-00&1.7520e-07$\pm$0e-00\\
3.562e-01&1.4852e-04$\pm$0e-00&8.1202e-06$\pm$0e-00&1.0313e-05$\pm$0e-00&8.1501e-06$\pm$0e-00&9.9982e-01$\pm$0e-00&1.7474e-07$\pm$0e-00\\
9.900e-01&1.6921e-04$\pm$0e-00&9.4711e-07$\pm$0e-00&7.6533e-08$\pm$0e-00&5.0200e-06$\pm$0e-00&9.9982e-01$\pm$0e-00&1.7437e-07$\pm$0e-00\\
      \hline
    \end{tabular}
  \end{center}
  \caption{Same as table A15 but for energy $E_{\rm in}=$1 TeV.} 
\end{table*}

\end{document}